\documentclass[journal=jacsat,manuscript=article]{achemso}

\usepackage[version=3]{mhchem} 

\usepackage[utf8]{inputenc} 
\usepackage[italian,english]{babel} 
\usepackage[T1]{fontenc} 
\usepackage{hyphenat} 
\usepackage{notes2bib} 
\usepackage{bibentry} 
\usepackage[inter-unit-product={}{-}{}]{siunitx} 
\usepackage{hyperref} 
\usepackage[shortlabels]{enumitem} 
\usepackage{graphicx} 
\usepackage{nicefrac} 

\author{Anna Maria Aloisi}
\email{annamaria.aloisi@istruzione.it}
\affiliation[Unknown University]
{già docente presso IPSIA A. Meucci, Cagliari (Italy)}
\altaffiliation{Via Tempio, 29 - 09127 Cagliari (Italy)}
\author{Pier Franco Nali}
\affiliation[Unknown University]
{Regione Sardegna, Cagliari (Italy)}
\altaffiliation{Via Tempio, 29 - 09127 Cagliari (Italy)}
\email{ampfn@tiscali.it}

\title[An \textsf{achemso} demo]
  {"Perché il mare è blu?": una rivisitazione in chiave pedagogica\footnote{First draft.}\\ \itshape "Why is the sea blue": a pedagogical review}

\abbreviations{IR,NMR,UV}

\begin{document}
\selectlanguage{italian}
\begin{abstract}
Le spiegazioni del colore del mare proposte a scopo didattico o divulgativo si rivelano spesso incomplete o eccessivamente semplificate, e dunque inadeguate alla comprensione della totalità del fenomeno. In questo articolo, dopo una rassegna storica delle indagini sull'argomento negli ultimi due secoli e un accenno allo stato attuale delle conoscenze, il problema viene affrontato in modo più rigoroso adottando un punto di vista comprensivo e pluridisciplinare. Viene proposto un modello fenomenologico in grado di fornire risultati operativi in contesto didattico. 
\end{abstract}
\selectlanguage{english}
\begin{abstract}
\textit{Popularized explanations of the color of the sea often prove to be incomplete or oversimplified, and hence inadequate to become acquainted with the phenomenon in its whole. In this paper, after a historical review of the investigations on the subject in recent centuries and a brief account of the current state of knowledge, the topic is addressed in a more rigorous way adopting a comprehensive and multidisciplinary viewpoint. A phenomenological model providing operational results in educational context is proposed.} 
\end{abstract}
\selectlanguage{italian}

\section{Introduzione}
Qualche tempo dopo la pubblicazione sull'ormai dismesso sito online della SISSA, {\url{http://ulisse.sissa.it}}, di un nostro saggio divulgativo sul problema dell'azzurro del cielo\cite{aloisienigma}, venne spontaneo a qualche lettore esigente chiederci di approfondire una questione che in quel testo avevamo accennato solo di sfuggita, riassunta nella domanda: "perché il mare è blu?"\cite{costanzo}. La domanda, apparentemente scontata poiché fa parte della nostra quotidianità, è ricca di spunti scientifici, ma può nascondere anche dei tranelli. La si incontra, con una certa frequenza, in testi scolatici, articoli e libri di divulgazione scientifica\cite{scarpa2004arcobaleno}, nonché in siti internet. Benché non manchino fortunatamente trattazioni eccellenti, sia su testi a stampa\cite{bohren2001clouds} che online\cite{oceanopticsbook}, il grado di comprensione del problema che mediamente si ricava da questo tipo di fonti è piuttosto insoddisfacente. Questo è dovuto soprattutto all'introduzione di semplificazioni eccessive, che ingenerano nel lettore un'impressione fallace di conoscenza o certezza, e a qualche inciampo teorico in cui incorrono persino libri di testo solitamente attendibili\cite{monk2004physical}. A loro parziale scusante va riconosciuto che la risposta alla apparentemente futile domanda è tutt'altro che banale: entrano in gioco complicati meccanismi chimico-fisici e di feedback (più che nel problema dell'azzurro del cielo), che non si prestano a essere decodificati in un unico concetto o in una formula onnicomprensiva. 

Le spiegazioni al problema del colore del mare proposte più di frequente chiamano in causa, di volta in volta: (1) l'assorbimento selettivo della luce da parte delle molecole d’acqua (cfr. ad es. rif.~\citenum{aloisienigma}), che essendo massimo all’estremità rossa dello spettro visibile rende il mare trasparente alla componente blu-violetta della radiazione solare; (2) la diffusione (o scattering) di Rayleigh (cfr. ad es. rif.~\citenum{scarpa2004arcobaleno}, p.~128, o anche rif.~\citenum{aloisienigma}), che avviene nei liquidi con un meccanismo simile a quello dell'azzurro del cielo; (3) a volte, l'effetto Raman (cfr. ad es. rif.~\citenum{monk2004physical}, p.~483), presente nei liquidi con un ruolo secondario nell'interpretazione del fenomeno del colore del mare. Anche la riflessione (di Fresnel) della luce dal cielo, benché da lungo tempo screditata come spiegazione principale del blu del mare, incontra ancora molti sostenitori nei blog online.

Ciascuna di queste spiegazioni -- o teorie -- è a suo modo corretta, ma parziale: è l'\textbf{\textit{attenuazione}} del fascio luminoso nell'acqua, costituita dalla combinazione dei due fenomeni dell'assorbimento e della diffusione, che rende conto, con ottima approssimazione, del colore blu del mare. (Non consideriamo l'emissione di radiazione termica dal mare, trascurabile nella regione del visibile). Questo inscindibile binomio "\texttt{attenuazione (c) = assorbimento (a) + scattering (b)}" viene messo in luce di rado o, per meglio dire, raramente viene espressa in modo chiaro la necessità che siano contemporaneamente presenti entrambi gli effetti (a) e (b) affinché il mare appaia di quel colore. Ancor più di rado questi due processi vengono descritti in modo adeguato alla complessità dell'ambiente marino, sottacendo che una descrizione realistica della fenomenologia osservata, sia in termini qualitativi che quantitativi, richiederebbe l'applicazione della teoria dell'interazione della luce nei mezzi assorbenti--diffondenti in presenza di una miriade di costituenti "in-acqua" (ioni salini dissolti, fitoplancton, varie specie di materiali inorganici e organici dissolti o in sospensione allo stato particellare, le bolle d'aria, ecc.) e di svariate condizioni esterne (atmosfera, irradiazione solare, fondale, punti di osservazione, ecc.) che influenzano il colore (e molte altre proprietà ottiche) dell'acqua di mare. 

Nel seguito tenteremo di ricostruire un quadro d'insieme coerente dei processi che entrano in gioco nella corretta (e completa) teoria del colore del mare, cercando di coniugare semplicità e rigore. Seguiremo un approccio pedagogico che potremmo definire "cartesiano", nel senso che individua nell'analisi e nella sintesi le operazioni fondamentali che portano alla conoscenza: dalla scomposizione del problema nelle sue componenti elementari alla ricomposizione dei processi analizzati mettendoli in relazione secondo i rapporti identificati dall'analisi. Questo approccio fornisce anche lo spunto per ricchi approfondimenti pluri-- e inter--disciplinari tra chimica, fisica, scienze naturali. Se ci si accontenta di limitare l'analisi a una dimensione puramente qualitativa è sufficiente un modesto apparato matematico, mentre un'analisi rigorosa non può prescindere dal complesso paradigma teorico - di derivazione astrofisica - del trasferimento radiativo, applicato al caso speciale della luce nel mare. Questo metodo consiste nello studio particolareggiato del trasferimento dell'energia raggiante attraverso la superficie aria--acqua e attraverso gli strati interni dei corpi d'acqua naturali, e permette di mostrare nel dettaglio i contributi dei singoli costituenti chimico-fisici nelle interazioni con i fotoni. Una trattazione di questo tipo andrebbe ben oltre i nostri scopi\cite{preisendorfer1976hydrologic,*mobley1994light}. Partiremo invece da una rassegna storica del dibattito plurisecolare che ha condotto infine all'interpretazione moderna del fenomeno. Le travagliate vicende storiche di questa lunga ricerca fanno anche intravedere come il metodo scientifico operi la selezione delle teorie e come non sempre la scienza segua un percorso lineare.
\section{Rassegna storica}
Tralasciando le ricerche più antiche\cite{WERNAND201335,*clarityonthesea}, il problema del colore del mare (e delle grandi distese d’acqua in genere) era stato studiato nella seconda metà dell’Ottocento da scienziati come J. Tyndall, J. Aitken e il chimico belga Walthère Spring (in particolare quest’ultimo aveva dedicato all’argomento lunghe e minuziose ricerche), che ne avevano ben compreso la complessità. Secondo la teoria a quel tempo accreditata il blu del mare era dovuto all’assorbimento selettivo da parte dell'acqua delle diverse lunghezze d'onda che compongono lo spettro del visibile, in combinazione con la diffusione della luce da particelle in sospensione (cfr. rif.~\citenum{bohren2001clouds}, p.~381). Come vedremo tra breve, questa teoria si avvicina molto all'interpretazione moderna ma manca ancora di alcune idee essenziali per completare il mosaico, prima fra tutte: una chiara consapevolezza della struttura molecolare della materia e delle proprietà ottiche delle molecole.
\subsection{R. Bunsen (1847)}
Bunsen \cite{bunsen1847ueber,*bunsen1847blaue,*bunsen1849colour} viene accreditato come il primo studioso a negare che l'acqua fosse incolore\cite{bancroft1919color}. Dalla \textit{Jahresbericht \"uber die Fortschritte der Chemie und verwandter Theile anderer Wissenschaften} (Relazione annuale sui progressi della chimica e parti correlate di altre scienze) del 1847 (rif.~\mciteSubRef{bunsen1847blaue}) apprendiamo che:

"\textit{Nelle pozze bianche delle sorgenti calde dell'Islanda, che sono rivestite con incrostazioni di ghiaia bianca, l'acqua è blu verdastra. Secondo Bunsen (rif.~\mciteSubRef{bunsen1847ueber}), l'acqua pura è blu, e le deviazioni da questo colore sono sempre dovute alle mescolanze o al riflesso di uno sfondo scuro o colorato. Lo si vede guardando oggetti bianchi lucidi su un fondo bianco attraverso uno strato d'acqua spesso 2 metri, contenuto in un tubo internamente annerito, o illuminato solo dalla luce solare che ha attraversato tale strato}". 

Dunque gli altri colori oltre al blu sono dovuti a sostanze estranee o alla riflessione di luce da uno sfondo colorato. Bunsen non si addentrò in dettagli sui meccanismi che provocano i cambiamenti di colore (cfr. rif.~\citenum{bancroft1919color}, p.~459), ma già l'aver intuito che l'acqua ha un colore intrinseco è un risultato notevole. Va notato, tuttavia, che l'esperimento con tubi anneriti, riprodotto nel 2003 da Wernand con tubi di poliuretano lunghi 4 metri riempiti con acqua milli-Q "ultrapura" (Type I, resistività $>$ \SI{18,2}{\mega\ohm\centi\meter} a \SI{25}{\celsius}), non fornisce un risultato chiaro come descritto da Bunsen: i tubi dovrebbero essere più lunghi per ottenere una colorazione blu evidente (cfr. rif.~\mciteSubRef{WERNAND201335}, p.~55 e p.~76).
\subsection{J. Tyndall (1870, 1871)}
Tyndall\cite{tyndall1869iv,*tyndall1870colour,*tyndall1871colour,*tyndall1871causes} compì esperimenti sul colore del mare durante il viaggio di ritorno dalla (fallita) spedizione in Algeria per l'eclissi del 22 dicembre 1870\cite{tyndall1886fragments}. Presentò i suoi risultati il 20 gennaio 1871 in un discorso alla Royal Institution (rif.~\mciteSubRef{tyndall1871colour}) che riguardava due argomenti apparentemente non collegati: il colore del mare e il rifornimento idrico di Londra. Tyndall esordì ricordando al pubblico "\textit{il fatto ben noto che un fascio di luce bianca inviato attraverso una camera oscura è visibile perché le piccole particelle fluttuanti di materia solida sono in grado di riflettere la luce dalle loro superficie e disperderla in tutte le direzioni. Se queste particelle venissero rimosse dall'aria, il pennello di luce non sarebbe visibile. Esattamente gli stessi fenomeni - continuò - sono prodotti dalle particelle solide tenute in sospensione in acqua.}" (rif.~\mciteSubRef{WERNAND201335}, p.~48). Dopo questa introduzione raccontò della fallita spedizione oltremare e di come aveva raccolto ed esaminato i campioni, così condividendo infine le sue conclusioni: "\textit{Oggi è risaputo che quando la luce bianca penetra nel mare i raggi rossi vengono attenuati per primi, poi successivamente i raggi arancione, giallo e verde e così via. Se non ci fossero particelle nel mare a retro-diffondere i raggi non assorbiti il mare sembrerebbe nero come l'inchiostro.}" (rif.~\mciteSubRef{WERNAND201335}, p.~49).

All'epoca lo scienziato irlandese è inconsapevole del fatto, dimostrato successivamente, che anche lo scattering molecolare svolge un ruolo importante e perciò l'acqua -- anche eliminando ogni residua particella in sospensione -- avrebbe ancora un aspetto blu scuro (indaco). 

Tyndall passò poi a descrivere l'esperimento del piatto da pranzo, che aveva compiuto durante il viaggio di ritorno in nave. "\textit{Da un lato della nave l'assistente di Tyndall, il signor Thorogood, calò un piatto di porcellana bianca con un peso di piombo fissato saldamente ad esso. Cinquanta o sessanta iarde di robusto filo di canapa erano attaccate al piatto. Dalla poppa, Tyndall osservò il piatto sommerso e vide a profondità considerevole la tonalità verde del piatto nell'indaco circostante.}" (rif.~\mciteSubRef{WERNAND201335}, p.~49). Immaginò che se il piatto bianco, che appariva come un oggetto verde quando veniva rimorchiato sott'acqua, fosse stato macinato in polvere e disperso, la porzione in sospensione di questa polvere avrebbe restituito un riflesso verde generale. Con questo esperimento intendeva provare la sua teoria sull'attenuazione della luce da parte dell'acqua di mare (per i dettagli di questa argomentazione cfr. rif.~\mciteSubRef{tyndall1871causes}). 

L'intervento di Tyndall ebbe grande risalto sulla stampa (non solo britannica) dell'epoca\cite{tyndall_1871Sydney}; più che altro - a onor del vero - per la parte finale dedicata al problema del rifornimento idrico di Londra, in quanto "soggetto di interesse per il lettore"\cite{tyndall_1871Engineer}. Il suo suggerimento di utilizzare per gli usi domestici l'acqua particolarmente pura ceduta dalle formazioni di gesso inglesi, rimuovendone artificialmente la durezza, originò una corrispondenza sul \textit{Times}\cite{tyndall_1871Times} in cui Tyndall sostenne la necessità di sperimentare la fornitura alla metropoli inglese di questa "acqua di gesso" -- anche se poteva soddisfare solo una parte della domanda di Londra -- poiché sarebbe stata molto più sicura. 

John Tyndall fu una figura centrale della scienza e della società vittoriana\cite{jackson2018ascent}. Personalità complessa e multiforme (fisico, inventore, ingegnere ferroviario, comunicatore di scienza, polemista, intrepido scalatore, profondo amante della poesia, \ldots), lo scienziato irlandese è ricordato come precursore degli studi sull'effetto serra e il riscaldamento globale, ma i suoi interessi scientifici spaziarono dal clima al magnetismo, dall'acustica alla batteriologia, e in molti altri campi. Per le sue indagini sull'azzurro del cielo (cfr. rif.~\mciteSubRef{tyndall1869iv}) Tyndall è anche considerato l'antesignano dell'approccio alla scienza come pura forma di scoperta, o \textit{curiosity-driven research}, che di lì ha preso anche il nome di \textit{blue-skies research}. Eppure, quale che fosse il suo punto di partenza, le sue ricerche ebbero anche importantissime ricadute pratiche finendo per fornire supporto alla teoria, a quei tempi controversa, che i microbi dell'aria sono responsabili delle malattie\cite{o'connell_2000}.
\subsection{J. Aitken (1882)}
Aitken\cite{aitken18821,*aitken1899colour} si propose di determinare quale fra le teorie a quel tempo in auge fornisse la corretta spiegazione della colorazione blu dell'acqua. Secondo una prima teoria, quella  della "riflessione selettiva", il colore è dovuto alla luce \textit{riflessa} (noi oggi diremmo \textit{diffusa}) da particelle materiali estremamente piccole sospese nell'acqua, così piccole che possono riflettere (diffondere) la luce solo alle piccole lunghezze d'onda, quelle appartenenti all'estremità blu dello spettro visibile. L'altra teoria, dell'"assorbimento selettivo", spiega il colore supponendo che l'acqua assorba selettivamente i raggi luminosi all'estremità rossa dello spettro visibile, cioè che l'acqua sia in effetti un mezzo trasparente al blu.

Per testare la correttezza di queste teorie rivali Aitken impiegò varie tecniche sperimentali, compreso un esperimento quasi identico a quello di Bunsen. Poté concludere che l'acqua assorbe selettivamente la radiazione all'estremità rossa dello spettro ma questa teoria dell'assorbimento selettivo non è sufficiente per render conto di tutti i differenti fenomeni osservati nell'acqua. Ma ciò che qui vogliamo soprattutto rimarcare è una considerazione "di metodo" che Aitken espone nel suo lavoro del \citeyear{aitken18821}:

"\textit{La causa del colore dell'acqua è stata un argomento frequente di speculazione. Ogni sostanza che è stata scoperta nell'acqua è stata di volta in volta indicata come causa del colore. Quando non è stato possibile trovare uno scopo utile per la sua presenza, le è stata attribuita una funzione ornamentale, per rendere l'acqua bella alla vista. Tutte queste speculazioni presumono che il colore sia dovuto a qualche impurità presente nell'acqua. Questo, però, significa ovviamente eludere l'intera questione. Per prima cosa è necessario sapere se l'acqua ha un colore in sé, e qual è quel colore, prima di poter dire qualcosa sull'effetto delle impurità.}" (rif.~\mciteSubRef{aitken18821}, p.~480).
\subsection{W. V. Spring (1883)}
Spring\bibnote{Spring dedicò una serie di pubblicazioni nell'arco di vari anni al colore delle acque marine e lacustri. La maggior parte delle opere di Spring sono accessibili online sul repository dell’università di Liegi, all'url \url{https://orbi.uliege.be/browse?type=authorulg&rpp=20&value=Spring\%2C+Walth\%C3\%A8re+p00022}. Un elenco di un centinaio di pubblicazioni si trova all'url \url{http://waltherespring1848-1911.e-monsite.com/pages/biographie-de-walthere-spring.html}. Le sue \textit{Oeuvres complètes} sono state pubblicate dalla Société Chimique de Belgique in 2 volumi (Bruxelles, 1914–1923) con una prefazione biografica tratta da Crismer L. \textit{Walthère Spring: sa vie et son œuvre}; Ad. Hoste, 1912.}, chimico e ingegnere minerario belga, ispirato dalle ricerche di Tyndall sul colore blu del cielo, riuscì, dopo un lungo e minuzioso lavoro, a osservare il "vero" colore dell'acqua naturale e di acque chimicamente pure. Utilizzando tecniche ingegnose e dimostrando estrema abilità di sperimentatore, fu tra i primi a preparare acqua otticamente "vuota", libera cioè da ogni traccia di particelle sospese, facendole depositare con l'aiuto di zinco precipitato colloidale e idrossido di alluminio formati dall'azione dell'ammoniaca sui sali solubili di zinco e alluminio\cite{lionetti1951walter}. 

Nei primi esperimenti con acqua distillata aveva ottenuto un blu puro come il blu del cielo, ma aveva notato che nel giro di pochi giorni l'acqua diventava verde-blu. Questo dimostrava che l'acqua distillata nelle condizioni di laboratorio non era perfettamente pura ma conteneva sostanze (che potevano essere minerali, organiche o anche organismi viventi) che cambiano nel tempo. Aggiungendo piccolissime quantità di bicloruro di mercurio (per non influenzare il colore intrinseco dell'acqua) osservò che la colorazione veniva preservata e l'acqua verde-blu tornava blu senza mostrare segni di cambiamento per settimane. Avendo aggiunto un prodotto chimico estremamente letale per i microorganismi (la parte organica) presenti nell'acqua distillata, Spring si convinse che il colore blu non era causato da particelle minuscole (la presenza di materiale inorganico poteva essere esclusa perché scompare dopo una corretta distillazione). Perciò - ragionava il chimico belga - doveva esserci nell'acqua qualcos'altro che causava il colore blu (\citeauthor{spring1883popular}, \citeyear{spring1883popular}, pp.~71--72). 
\begin{figure}
  \includegraphics[width=\textwidth]{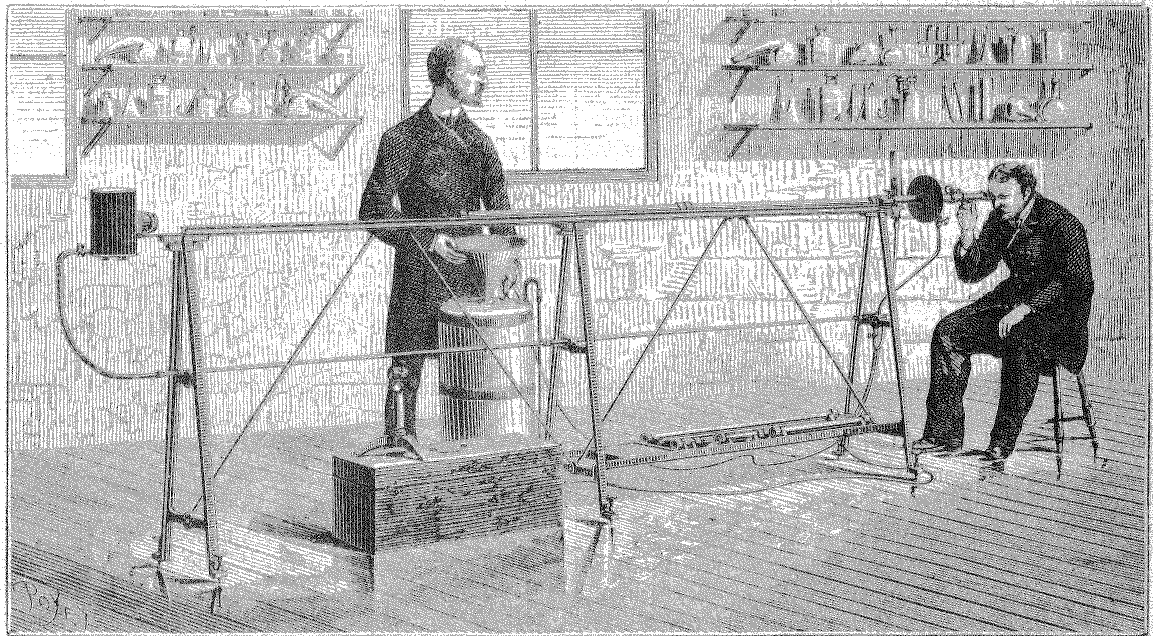}
  \caption{Uno spettroscopio d'assorbimento simile allo strumento utilizzato da Spring per investigare sul "vero" colore dell'acqua\cite{DeThierry1886LaNature}. Spring usò la luce solare invece della lampada elettrica visibile nella figura per illuminare la colonna d'acqua.}
  \label{fgr:graph01}
\end{figure}

Per rendere l'idea della difficoltà degli esperimenti condotti da Spring per osservare il colore intrinseco dell'acqua pura basti dire che l'acqua doveva essere contenuta in tubi di vetro di \SI{15}{\milli\meter} di diametro e lunghi fino a \SI{26}{\meter}, la cui sperimentazione richiedeva strumenti ottici di estrema precisione e una grande abilità manuale. La difficoltà risiedeva nel fabbricare un tubo di questa lunghezza coassiale con il raggio di luce che doveva attraversarlo. Furono necessarie quasi sei settimane di lavoro per ottenere l'allineamento dell'apparato. Dopo un duro lavoro, Spring riuscì a completare questi esperimenti di precisione e riferì che (anche quando le distanze attraversate erano di soli 4 o 5 metri) il colore naturale dell'acqua è "\textit{un blu ceruleo simile a quello del cielo allo zenit visto da una posizione elevata.}" (\citeauthor{spring1883popular}, \citeyear{spring1883popular}, p.~72). Il chimico belga determinò che l'acqua pura è blu mentre l'acqua ordinaria è verde o marrone in virtù delle particelle di acido umico che vi sono sospese. Estendendo le ricerche sul colore ai liquidi organici, trovò che gli idrocarburi hanno una colorazione gialla, e gli alcoli -- cosa piuttosto interessante -- verde, il colore intermedio tra l'acqua e le paraffine (cfr. rif.~\citenum{lionetti1951walter}, p.~605). Non riuscì tuttavia a rendere i liquidi organici otticamente vuoti, ossia privi del fenomeno di Tyndall, scopo che fu invece raggiunto in esperimenti contemporanei condotti in Italia\cite{battelli1899sull}.

Spring è più noto -- come già Tyndall -- come precursore degli studi sull'effetto serra\cite{demaree2009walthere}, ma la sua opera si caratterizzò per l'interesse verso problemi riguardanti l'intero arco dei fenomeni naturali che non avevano ancora ricevuto un'adeguata spiegazione, unita a originalità di approccio, ingegnosità sperimentale, abilità manuale, e a una chiarezza e forza espositiva non comuni. Anche la versatilità dei suoi interessi è particolarmente degna di nota\cite{gillispie1970dictionary}. 
\subsection{La situazione a fine Ottocento}
In sintesi possiamo dire che, grazie all'opera di questi pionieri (e di Soret, Sainte-Claire Deville, e altri, su cui sarebbe troppo lungo soffermarci), nella seconda metà dell'Ottocento si era cominciato a intravedere gradualmente il corretto ruolo dei vari fattori che influenzano il colore del mare\cite{plass1978color}. Particolarmente significativo è, a questo riguardo, il contributo dell'astrofisico italiano Annibale Riccò\cite{ricco1876studi,*ricco1879studi,*ricco1904sul}, il quale compì esperienze con i tubi, utilizzando lo spettroscopio per queste indagini, e ripeté gli esperimenti con il disco di Tyndall, giungendo a conclusioni molto vicine a quelle della teoria sull'attenuazione dello studioso irlandese. Particolare curioso, nel suo primo resoconto Riccò riporta un simpatico aneddoto su un marinaio siculo che accompagnava gli scienziati nelle loro esperienze. L'uomo, interrogato sulla differenza di colore dell'acqua del porto e in alto mare, rispose nel suo caratteristico dialetto: "\textit{l'acqua è virdi ccà, pirchì è trubula: in menzu a lu mari spurchizzi nun ci nni stannu e pri chistu l'acqua ddà è blù}" (rif.~\mciteSubRef{ricco1876studi}, p.~114). Nella terza memoria (rif.~\mciteSubRef{ricco1904sul}), Riccò menziona i ruoli complementari delle cosiddette \textit{\textbf{teoria chimica}} (o dell'\textbf{\textit{assorbimento}}) e \textbf{\textit{teoria fisica}} (ossia della \textbf{\textit{diffusione}}) -- con le quali si era confrontato Aitken e ancora dibattute nei primi anni del Novecento -- per spiegare la colorazione dell'acqua: la prima sarebbe valida nel regime delle grandi lunghezze d'onda e spiegherebbe l'intercettamento e l'eliminazione di questi raggi nella luce trasmessa dall'acqua come una caratteristica costituzionale e propria dell'acqua pura; la seconda varrebbe alle piccole lunghezze d'onda e spiegherebbe l'azione delle minutissime sostanze estranee sospese.
\subsection{L. Rayleigh (1910)}
Lord Rayleigh (J. W. Strutt) irruppe nel dibattito nel \citeyear{rayleigh1910colours}, sull'onda dell'interesse per il problema del colore del mare nato in seguito a un viaggio attorno all'Africa che aveva intrapreso nel 1908. Presentò le sue conclusioni in un discorso tenuto alla Royal Institution il 25 febbraio 1910\cite{rayleigh1910colours}, in cui affermò (correttamente) che non si può vedere il giusto colore dell'acqua se la luce del sole non attraversa uno spessore sufficiente prima di raggiungere l'osservatore. La profondità dell'oceano, naturalmente, è abbastanza grande da consentire che il colore si manifesti. Ma -- continua Rayleigh -- se l'acqua dell'oceano è molto pulita non c'è niente in essa che possa rispedire la luce verso un osservatore e perciò, in queste condizioni, non si può vedere il suo colore intrinseco e il blu scuro del mare profondo è   semplicemente “il riflesso del blu del cielo”. 

Quest'ultima -- errata -- conclusione può forse sorprendere per uno scienziato della levatura di Rayleigh, ma va notato che questi era ancora legato alla concezione ottocentesca secondo cui non è la massa d'acqua in sé a diffondere la luce ma è necessaria della materia sospesa "nell'acqua" per ridirigere la luce verso l'osservatore (a meno che non sia il fondale stesso a fungere allo scopo). Si ricordi, infatti, che la teoria molecolare della materia non era ancora consolidata all'epoca e vi erano motivi per dubitare che queste ipotetiche molecole si comportassero come le sfere dielettriche della teoria che aveva funzionato così bene per spiegare il blu del cielo. Rayleigh, inoltre, riteneva non del tutto attendibili le osservazioni degli sperimentatori ottocenteschi, avendo egli stesso tentato con scarso successo di riprodurre autonomamente alcuni loro risultati. Leggiamo infatti in un passo del suo discorso: "\textit{Alla luce di queste prove riesco a malapena a evitare la conclusione che il bluastro dell'acqua in lunghezze di 4 metri sia stato esagerato, in particolare da Spring, sebbene non abbia motivo di dubitare che un blu completamente sviluppato si possa ottenere con spessori molto maggiori.}" (rif.~\citenum{rayleigh1910colours}, p.~49). La sua ipotesi al riguardo era che gli sperimentatori non avessero posto sufficiente attenzione ad usare in partenza una sorgente che producesse effettivamente luce bianca.
\subsection{C. V. Raman (1922)}
Raman\cite{raman1922molecular} partì dalle osservazioni compiute in seguito a una traversata oceanica -- che, per inciso, gli diedero l'ispirazione per tutta la sua ricerca successiva\cite{raman1953molecular} -- e dall'accordo tra i risultati sperimentali e la teoria delle fluttuazioni di Einstein-Smoluchowski (e, conseguentemente, dall'inevitabilità delle conseguenze fisiche di questa teoria di scattering molecolare) per formulare una critica alla posizione di Rayleigh, riprendendo gli stessi argomenti che questi aveva portato a sostegno delle sue tesi e sovvertendone le conclusioni. La critica principale toccava il nucleo dell'argomentazione di Rayleigh, ossia l'asserita invisibilità del colore dell'acqua pura. Se infatti si postula un oceano di acqua pulita (totalmente priva di materia in sospensione), e molto profondo, si deve necessariamente accettare la conseguenza (derivante dalla teoria di Einstein-Smoluchowski) che esso diffonderebbe luce di un colore che supera di molto il blu del cielo in saturazione ed è di luminosità paragonabile (cfr. rif.~\citenum{raman1922molecular}, p.~71). 

Raman "smonta" uno per uno vari altri ragionamenti legati alla teoria di Rayleigh, e riserva infine alcune considerazioni all'esperimento di Tyndall del piatto sommerso, che è interessante riportare. Non si deve presumere che il colore apparente del piatto corrisponda a quello trasmesso da una colonna d'acqua pari al doppio della profondità (doppio perché la luce deve prima discendere e poi risalite la colonna d'acqua). L'effetto osservato è in realtà la risultante di due differenti fattori: (a) la luce diffusa verso l'alto dalle molecole d'acqua della colonna interposta tra il piatto e la superficie (luce che non ha raggiunto il piatto perché ha urtato una molecola interposta lungo il cammino); (b) la luce diffusa dal piatto che raggiunge l'osservatore dopo aver subito l'attenuazione passando attraverso la colonna d'acqua. L'effetto (b) è a sua volta composto: (c) dall'effetto della luce che raggiunge il piatto e viene da questo diffusa seguendo il regolare percorso della colonna; (d) dall'effetto dell'illuminazione laterale del piatto da parte della luce diffusa tutt'intorno dalle molecole del liquido. Il sommarsi di questi effetti fa sì che il piatto appare molto più blu di quanto lo sia realmente la luce trasmessa (cioè se non operasse il fenomeno dello scattering ma solo la trasmissione della luce in avanti). Ne segue che le osservazioni della luce trasmessa con questo metodo si rivelano incapaci di produrre risultati sperimentalmente utili in quanto non consentono di separare gli effetti della trasmissione da quelli dello scattering (cfr. rif.~\citenum{raman1922molecular}, p.~79).  

Possiamo così riepilogare alcune delle principali conclusioni del lavoro di Raman:
\begin{enumerate}[(a)]
\item viene proposta una nuova teoria del colore del mare, cioè che esso è dovuto allo scattering molecolare della luce in acqua;
\item lo scattering molecolare in acqua è una conseguenza necessaria della teoria delle fluttuazioni di Einstein-Smoluchowski ed è circa 160 volte più intenso che nell'aria purificata;
\item uno strato sufficientemente profondo di acqua pura appare per effetto dello scattering molecolare di un blu più saturo del blu del cielo e di intensità paragonabile; 
\item il colore è determinato in primo luogo dalla "diffrazione" (così nella terminologia del lavoro originale di Raman), l'assorbimento gli conferisce solo un tinta più satura;
\item viene mostrato che le precedenti teorie sul blu del mare dovuto alla riflessione del cielo o alla materia sospesa nell'acqua sono errate;
\item viene evidenziato che il colore osservato di un piatto bianco immerso a una certa profondità sotto acqua trasparente non corrisponde realmente al carattere della luce trasmessa (futilità dell'esperimento di Tyndall).
\end{enumerate}

Vale infine la pena sottolineare che nello scattering molecolare considerato nella teoria di Raman operano meccanismi di tipo elastico (cioè senza cambiamento di lunghezza d'onda), che vanno distinti dai processi di diffusione anelastica (con cambiamento di lunghezza d'onda, come l'omonimo effetto che sarà scoperto più tardi da Raman) che pure avvengono nei liquidi e nel mare stesso (cfr. rif.~\citenum{monk2004physical}, p.~483), come vedremo meglio più avanti.
\subsection{K. R. Ramanathan (1922, 1923, 1925)}
Il punto di vista di Raman fu confermato dal suo collaboratore Ramanathan\cite{ramanathan1922molecular,*ramanathan1923lviii,*ramanathan1925transparency} mediante esperimenti sullo scattering della luce in liquidi e vapori, tesi a verificare che esiste un legame fondamentale tra tutti questi fenomeni e che, in tutti questi casi, la formula di Einstein-Smoluchowski è in accordo con i risultati sperimentali (cfr. rif.~\mciteSubRef{ramanathan1923lviii}, p.~160). 

In una successiva investigazione, sempre utilizzando la formula di Einstein-Smoluchowski (in una variante corretta per uno scattering addizionale prevalentemente non-polarizzato, dovuto all'anisotropia delle molecole, onde tener conto del fatto sperimentale che la luce diffusa trasversalmente non è perfettamente polarizzata) mostrò che un oceano ideale di acqua pura, per gli effetti combinati di scattering molecolare e assorbimento, sarebbe di colore blu-indaco. La presenza di \textit{piccole} quantità di materia sospesa non influirebbe apprezzabilmente sul colore. Con l'aumento di queste quantità il colore cambierebbe via via a bluastro, verde, bianco-verdastro e bianco (cfr. rif.~\mciteSubRef{ramanathan1923lviii}, p.~552). 
\subsection{W. V. Shoulejkin (1923, 1924)}
Shoulejkin\cite{shoulejkin1923color,*shoulejkin1924color} sembra sia stato il primo, nel \citeyear{shoulejkin1923color}, a combinare le idee in circolazione e a sviluppare una spiegazione completa del colore del mare\cite{dickey2011shedding}, deducendo una formula che tiene conto di tutti i fattori che intervengono nella produzione del colore: la diffusione della luce da parte di molecole d'acqua e particelle più grandi (bolle d'aria, sostanze sospese, \ldots) e l'assorbimento da parte di molecole e sostanze dissolte\cite{jerlov1976marine}, nonché il riflesso della luce dal cielo. Egli è inoltre considerato l'iniziatore dello studio teorico del trasferimento radiativo nel mare, campo nel quale dimostrò il fatto notevole che la distribuzione della radianza tende a uno stato stazionario con l'aumentare della profondità del mare.\cite{shoulejkin1933data}

Shoulejkin notò che il colore del mare (e dei laghi) non può essere spiegato soltanto con:
\begin{enumerate}[(1)]
\item il colore intrinseco dell'acqua dovuto all'assorbimento selettivo; \\ oppure
\item lo scattering (di Rayleigh) da particelle microscopiche sospese o bolle di gas.
\\ Sembra necessaria una combinazione di queste due cause, insieme con:
\item la "riflessione" (nella terminologia di S., equivalente alla riflessione diffusa o scattering di Mie) selettiva da parte di particelle più grandi (polvere fine, plancton, ecc.); 
\\ e inoltre con
\item la riflessione della luce dal cielo (quando si considera anche l'effetto delle onde).
\end{enumerate}
Sulla base di una formula relativamente semplice lo studioso sovietico riesce a dimostrare matematicamente che, in assenza di diffusione da parte di particelle sospese, si otterrebbe una superficie perfettamente nera; se viceversa ad essere inibito fosse l'assorbimento selettivo, la superficie del mare sarebbe perfettamente bianca (cfr. rif.~\mciteSubRef{shoulejkin1923color}, p.~88). 

È da notare che Shoulejkin non considera, diversamente da Raman, lo scattering molecolare "vero" dell'acqua, ma concentra l'analisi su particelle sospese nell'acqua, che vengono distinte in due tipi (o ordini). Le particelle del primo tipo sono così piccole che diffondono la luce secondo la legge di Rayleigh. Ma se insieme a queste particelle vi sono in acqua anche particelle più grandi (del secondo tipo), che causano una "riflessione" selettiva (terminologia di S., approssimativamente corrispondente alla riflessione diffusa o diffusione di Mie) e un assorbimento selettivo della luce, la formula della distribuzione spettrale della luce diventa più complicata (cfr. rif.~\mciteSubRef{shoulejkin1924color}, p.~745). Lo scattering molecolare considerato da Raman e Ramanathan è, secondo Shoulejkin, solo una componente molto piccola e trascurabile dello scattering della luce nel suo complesso, che è causato principalmente da agenti più energici. Nel Mar Nero, il più trasparente tra i mari studiati dal nostro, il rapporto tra la diffusione totale e la diffusione molecolare (calcolata) è circa 14:1 (cfr. rif.~\mciteSubRef{shoulejkin1924color}, p.~750). 

Un altro punto di disaccordo con Ramanathan riguarda la questione dell'influenza delle particelle di materia sospesa nell'acqua sul colore del mare, che il secondo riteneva trascurabile. Shoulejkin sostenne invece di aver mostrato, sperimentalmente, l'esistenza dell'effetto anche con particelle molto più grossolane di quelle presenti nell'acqua di mare (cfr. rif.~\mciteSubRef{shoulejkin1924color}, p.~750, e rif.~\mciteSubRef{shoulejkin1923color}, p.~97). 
\begin{figure}
\includegraphics[width=\textwidth]{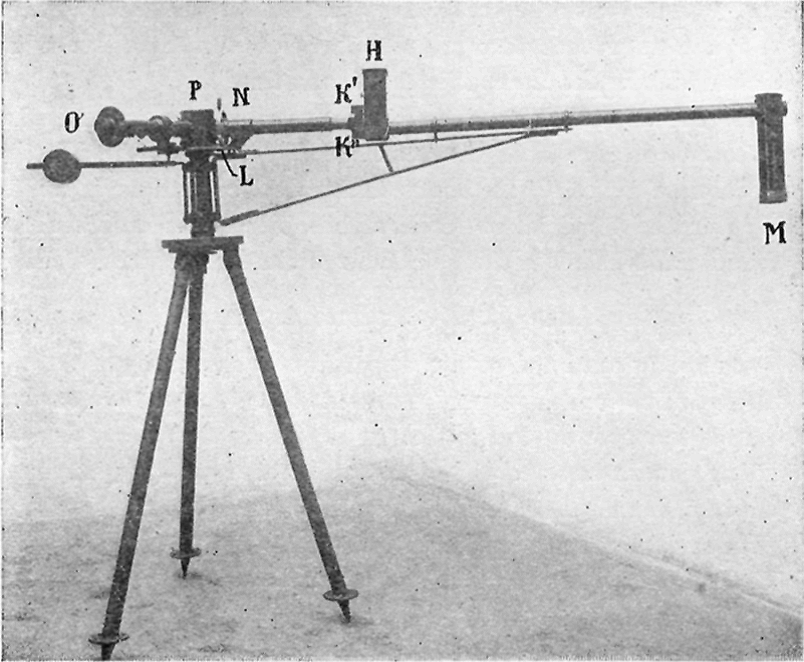}
\caption{Spettrofotometro marino di Shoulejkin. Forse, il primo dispositivo di telerilevamento per la misurazione del colore dell'acqua di mare. L'obiettivo \textit{M} si protende fuori bordo per misurare la radianza di risalita. L'obiettivo \textit{H} misura la luce del sole e del cielo entrante nello strumento. La luce viene trasmessa verso la fessura K" di un doppio collimatore mediante prismi e lenti. I due spettri sovrapposti venivano osservati attraverso l'oculare \textit{O}. (Fonte:~\citeauthor{phdthesis},~\citeyear{phdthesis}).}
\label{fgr:graph02}
\end{figure}

Ne seguirà un dibattito, con successivo intervento dello stesso Ramanathan (cfr. rif.~\mciteSubRef{ramanathan1925transparency}), il quale mostrerà come, contrariamente all'opinione di Shoulejkin, la \textit{luminosità} dovuta al solo scattering molecolare sia apprezzabile, pari a circa un sesto di quella del cielo blu. 

Guardando retrospettivamente, si può accettare l'osservazione di Shoulejkin sulla prevalenza dello scattering causato dalla materia in sospensione nelle acque da lui esaminate, non foss'altro che per il fatto che nella maggior parte delle acque marine oltre alle molecole d'acqua è presente materiale in sospensione. Le eccezioni sono acque estremamente oligotrofiche che si trovano molto lontano dalle coste e sono paragonabili ad acque della massima purezza, in cui effettivamente solo le molecole sono responsabili del colore dell'acqua\cite{wernand2011ocean}. 

Meno condivisibile il suo suggerimento che il colore di queste acque pure è spiegato dal colore del cielo riflesso dalla superficie del mare (cfr. rif.~\mciteSubRef{shoulejkin1924color}, p.~751). La riflessione gioca infatti un ruolo, non trascurabile, ma relativamente debole rispetto alla luce diffusa verso l'alto dalla massa d'acqua\cite{cabannes1949azul}. In ogni caso, è facile vedere che il colore del mare non si può spiegare interamente con la riflessione della luce dal cielo. Lo dimostrano le osservazioni sulla polarizzazione della luce proveniente dal mare. Se questa fosse interamente prodotta da riflessione speculare dalla superficie marina dovrebbe essere in parte polarizzata parallelamente alla superficie stessa con un massimo quando l’angolo d’incidenza è pari all'angolo di Brewster (\ang{53} per l'acqua). Usando un polarizzatore e ruotandolo opportunamente, in modo che non lasci passare la polarizzazione orizzontale, si possono eliminare i riflessi, ma nondimeno si continua a osservare radiazione non polarizzata proveniente dal mare. 

Comunque sia, la teoria di Shoulejkin, pur con alcuni limiti e nella sua relativa semplicità, si rivela utile e conduce a conclusioni operative. Ne faremo tesoro più avanti nell'illustrazione di un modello a scopo didattico.
\subsection{J. Lenoble (1956)}
Non possiamo chiudere questa nostra carrellata senza un breve accenno, tra tanti uomini, all'importante lavoro di una donna, M.lle Jacqueline Lenoble\cite{lenoble1956remarque}, a metà degli anni cinquanta del Novecento. Applicando il metodo di Chandrasekhar del trasferimento radiativo\cite{chandrasekhar1950radiative} al calcolo del colore del mare, M.lle Lenoble dimostra teoricamente che un ipotetico mare d'acqua distillata con una diffusione non selettiva (indipendente dalla lunghezza d'onda), considerato infinito e illuminato da un cielo uniformemente bianco, sarebbe blu. L'applicazione del modello a diversi casi reali mostra che nelle acque limpide la lunghezza d'onda dominante si colloca vicino ai \SI{480}{\nano\meter}. 
Il colore del mare si può spiegare sia con l'assorbimento selettivo del rosso, sia con la diffusione selettiva del blu. La prima spiegazione\cite{legrand1939penetr} sembrerebbe \textit{a priori} più soddisfacente, poiché, nel mare, la diffusione molecolare, inversamente proporzionale alla quarta potenza della lunghezza d'onda, è debole rispetto alla diffusione poco selettiva delle particelle più grandi, mentre il minimo di assorbimento nel blu è nettamente marcato (cfr. rif.~\citenum{lenoble1956remarque}, p.~663). Infine, l'assorbimento selettivo dell'acqua permette di spiegare il blu del mare senza far intervenire la diffusione selettiva (cfr. rif.~\citenum{lenoble1956remarque}, p.~664).
\subsection{Cenno alle ricerche moderne}
In tempi più recenti si è assistito a un crescente sviluppo delle tecniche sperimentali di misura del colore dell'oceano e dei modelli teorici e computazionali per calcolare in dettaglio i contributi relativi dei diversi fattori che lo influenzano. Le misure del colore sono state fatte in numerose aree distinte dell'oceano globale insieme con la quantità di clorofilla presente (che ha un caratteristico profilo spettrale), non solo alla superficie ma in profondità. Particolare rilievo hanno assunto anche le tecniche di telerilevamento del colore da satellite. Nello specifico, tali tecniche consentono di stimare la biomassa dalla variazione del colore in funzione della concentrazione della clorofilla. Gli strumenti teorici comprendono oggi raffinati modelli di simulazione numerica (HydroLight, Monte Carlo, ecc.) i quali, includendo tutti i fattori chimico--fisici conosciuti e tutti i dati rilevati, permettono di seguire i fotoni attraverso le loro interazioni nell'atmosfera e nell'oceano: vengono considerati nei calcoli tutti gli ordini di scattering che danno un contributo apprezzabile al risultato. Questi metodi sono utilizzati per calcolare il trasferimento radiativo nel sistema oceano--atmosfera, la polarizzazione della radiazione, l'effetto delle onde sulla superficie marina, e molto altro.
\begin{figure}
\includegraphics[width=\textwidth]{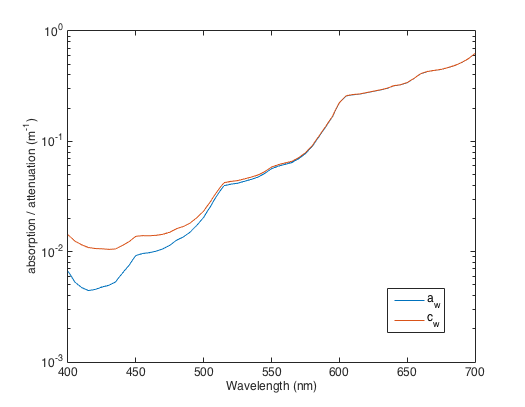}
\caption{Esempio di grafico semilogaritmico ottenuto con il programma HydroLight usando il data file \texttt{H2OabDefaults.txt}: \texttt{Default "pure water" absorption (a) and scattering (b) coefficients}. La curva blu rappresenta l'assorbimento selettivo $a_w$, la rossa l'attenuazione (\texttt{assorbimento + scattering}) $c_w=a_w+b_w$ in funzione della lunghezza d'onda della luce nell'acqua otticamente pura. I dati sono basati su:~\citeauthor{smith1981optical},~\citeyear{smith1981optical}. (Fonte: \texttt{oceanopticsbook.info})} 
\label{fgr:PureWaterACLogScale}
\end{figure}
\section{Discussione del fenomeno}
Applicando il metodo analitico-sintetico, il problema del colore del mare può essere scomposto analizzando i molti fattori che lo influenzano\cite{nielsen1974optical}. Per non appesantire troppo l'esposizione, qui consideriamo separatamente, ed esclusivamente su un piano qualitativo, i seguenti effetti principali:
\begin{enumerate}[(1)]
\item il colore intrinseco dell'acqua dovuto all'assorbimento selettivo;  
\item lo scattering molecolare elastico o anelastico;
\item lo scattering da particelle microscopiche sospese o bolle di gas; 
\end{enumerate}

Seguendo una terminologia mutuata dalla ricerca oceanografica più recente, distinguiamo preliminarmente le acque, cosiddette, di \textit{Caso 1} (mare aperto) dalle acque di \textit{Caso 2} (acque costiere)\cite{gordon2012remote}. Le acque di Caso 1, che rappresentano la parte di gran lunga preponderante dell'oceano globale, sono caratterizzate dall'influenza dei seguenti componenti (sempre in associazione tra loro) sulle proprietà ottiche dell'acqua del mare:
\begin{enumerate}[(1)]
\item fitoplancton in concentrazione variabile;
\item scarti (organici) associati (al fitoplancton), originati da pascolo dello zooplancton e da decadimento naturale;
\item sostanza organica dissolta, liberata da alghe e loro scarti (sostanza gialla, detta anche \textit{chromophoric dissolved organic material -- CDOM}).
\end{enumerate}
Le acque di Caso 2 possono contenere (o meno) i componenti 1, 2 e 3. Esse si differenziano dalle acque di Caso 1 per almeno una delle seguenti caratteristiche:
\begin{enumerate}[(1)]
\setcounter{enumi}{3}
\item la loro elevata torbidità (carico sedimentario) dovuta a sedimenti rimossi dal fondale lungo la linea di costa e nelle aree poco profonde e/o a particelle terrigene trasportate da fiumi e ghiacciai (queste si dicono acque di Caso 2 dominate da sedimenti);
\item elevato contenuto di sostanza organica dissolta drenata dal terreno (sostanza gialla terrigena), che le caratterizza come acque di Caso 2 dominate da sostanza gialla;
\item l'influenza cumulata dei fattori precedenti, cui può aggiungersi l'influsso antropico (sorgenti urbane e attività industriali che possono produrre sostanze particolate e sostanze dissolte).
\end{enumerate}
Altri fattori, quali polveri meteoriche e zooplancton, hanno un'influenza trascurabile sulle proprietà ottiche e li possiamo tralasciare.

Semplificando un po', si può dire che le acque di Caso 2 dominate da sedimenti mostrano uno scattering relativamente alto, a differenza di quelle dominate da sostanza gialla, con scattering relativamente più basso (cfr. rif.~\citenum{gordon2012remote}, pp.~29--30). Nelle acque più pure di Caso 1, con poco particolato, le proprietà ottiche dipendono principalmente dall'assorbimento e dallo scattering delle molecole d'acqua. Acque di questo tipo sono chiamate "hydrosol". Ci concentreremo pertanto, nel seguito, principalmente su questa casistica, che è largamente prevalente e ammette una trattazione relativamente semplice. Partiamo allora dal primo termine del nostro binomio "\texttt{c = a + b}": l'assorbimento selettivo della luce da parte della massa d'acqua (\texttt{a}).
\subsection{L'assorbimento selettivo della luce nel mare}
L’assorbimento selettivo della luce nell’acqua pura ha un minimo intorno a \SI{420}{\nano\meter} nella regione blu dello spettro visivo (cfr. rif.~\citenum{dickey2011shedding}, fig.~1, p.~44). Questo minimo deriva in via principale dalla bassa densità degli stati corrispondenti ai modi vibrazionali ad alta frequenza dei legami \ce{O\bond{-}H} delle molecole d'acqua, richiesti per l'assorbimento nel blu, dal momento che i modi vibrazionali fondamentali avvengono nell'infrarosso a circa \SI{3}{\micro\meter}\bibnote{La letteratura sull'argomento è davvero sterminata. Per un primo riferimento si possono esplorare le seguenti pagine di Wikipedia: \url{https://en.wikipedia.org/wiki/Electromagnetic_absorption_by_water}, \url{https://en.wikipedia.org/wiki/Molecular_vibration}.}. Si tratta di una caratteristica peculiare dell'acqua pura. I modi rotazionali e le interazioni elettroniche dirette, dominanti in quasi tutte le altre sostanze, hanno un effetto secondario sul colore dell'acqua. Vi sono poi un certo numero di altri effetti -- come vibrazioni di legami idrogeno intermolecolari -- che intervengono nelle fasi condensate dell'acqua a causa dell'interazione di strutture supermolecolari con la luce\cite{wozniak2007light,*B200372D,*braun1993water}. Naturalmente, l'acqua pura fa la parte del leone rispetto alle altre sostanze nell'assorbimento della luce nel mare, per l'assoluta superiorità numerica delle sue molecole nell'oceano globale (che si calcola ne costituiscano il 97\% del totale); tuttavia, nel mare è presente praticamente ogni elemento naturale\cite{pilson2013introduction}, sia pure quasi sempre in concentrazione molto bassa, ciascuno capace di fornire un proprio contributo all'assorbimento della luce.

\begin{table}
  \caption{Concentrazioni dei principali costituenti \\ ionici delle acque marine superficiali}
  \label{tbl:tab1}
  \begin{tabular}{ll}
    \hline
    Componente  & \SI{}{\gram/\kilogram} di acqua \\ & (sal. 35\textperthousand, temp. \SI{20}{\celsius}) \\
    \hline
    Ione cloruro (\ce{Cl^-}) & 19,353   \\
    Ione sodio (\ce{Na^+}) & 10,782  \\
    Ione solfato (\ce{SO_4^{2-}}) & 2,712 \\
    Ione magnesio (\ce{Mg^{2+}}) & 1,284 \\
    Ione \ldots & 0,\ldots \\
    \hline
  \end{tabular}
\end{table}
L'assorbimento della radiazione da parte di atomi, ioni salini, gas e altre sostanze inorganiche dissolte avviene attraverso meccanismi di eccitazione elettronica, ma l'effetto risulta molto piccolo nella regione visibile dello spettro ottico (può invece essere talvolta apprezzabile nell'infrarosso e nell'ultravioletto). I composti organici complessi (dissolti o contenuti nel plancton o nei rifiuti organici) costituiscono, dopo le molecole d'acqua, il principale agente responsabile dell'assorbimento della luce nell'acqua di mare. A causa della complessità delle molecole costituenti queste sostanze organiche, la soluzione delle equazioni quanto-meccaniche che descrivono queste molecole diventa, nonostante i progressi nella comprensione teorica e nelle tecniche computazionali, estremamente ardua e dispendiosa in termini di risorse di calcolo. Per semplificare, si è soliti assumere che le caratteristiche degli spettri di assorbimento si possano interpretare senza necessità di considerare la molecola complessa nella sua interezza, ma siano dovute a frammenti più o meno grandi di essa, noti come \textit{cromofori}. In altre parole, il cromoforo è il componente della molecola che conferisce il colore alla sostanza assorbendo alcune lunghezze d'onda della luce bianca che l'attraversa, facendone emergere luce di un colore complementare (tralasciamo il caso più complicato di più gruppi cromofori presenti in una stessa molecola). Il processo sottostante all'assorbimento è quello dell'eccitazione degli elettroni di valenza delle molecole. Sono molto comuni nel mare sostanze contenenti il cromoforo "blu-violetto" che, assorbendo appunto questi colori, conferisce alla "sostanza gialla" il caratteristico colore (cfr. rif.~\citenum{wozniak2007light}, p.~83). I pigmenti (sopratutto clorofilla) contenuti nelle cellule di fitoplancton marino sono il principale gruppo di sostanze organiche che assorbono luce nell'oceano (cfr. rif.~\citenum{wozniak2007light}, p.~295). Il meccanismo di assorbimento nel caso di molecole contenenti molti atomi come i pigmenti è assai più complesso che per l'acqua. Senza entrare troppo nei dettagli, possiamo dire che la presenza di modi vibrazionali (che cadono nell'infrarosso) permette un maggior numero di possibili transizioni elettroniche nelle lunghezze d'onda in vicinanza di transizioni elettroniche (nella regione visibile e ultravioletta) tra subshells. Queste transizioni addizionali, che includono sia eccitazioni elettroniche che vibrazionali, aggiungono righe spettrali ulteriori allo spettro di assorbimento, arricchendone la struttura. Se poi si considerano anche gli stati quantizzati associati ai modi rotazionali delle molecole (che richiedono un'energia molto piccola, nel range delle microonde), i livelli energetici permessi diventano molto densamente spaziati, tanto da apparire continui alla risoluzione della maggior parte degli strumenti. Considerazioni di questo genere possono spiegare in modo qualitativo come si originano spettri di assorbimento continui come quelli della clorofilla.

\begin{figure}
\includegraphics[width=\textwidth]{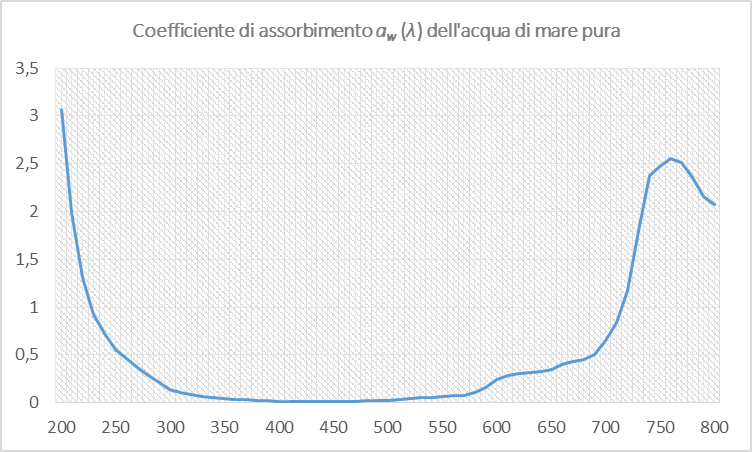}
\caption{Curva del coefficiente di assorbimento $a_w(\lambda)$ dell'acqua di mare pura (hydrosol), tracciata utilizzando i dati di \citeauthor{smith1981optical}, \citeyear{smith1981optical}. L'assorbimento è minimo in corrispondenza di $\lambda=\SI{430}{\nano\meter}$.}
\label{fgr:graphaw}
\end{figure}

In linea di principio, il coefficiente di assorbimento dei costituenti oceanici si può esprimere come somma dei contributi dovuti all'acqua e alle varie categorie di costituenti prima elencati: (1) fitoplancton, (2) materiale particolato, (3) sostanza organica dissolta (\textit{CDOM}). Come già rimarcato, l'assorbimento da parte dell'acqua è debole nel blu e intenso nel rosso e varia leggermente con la temperatura e la salinità. L'assorbimento del fitoplancton dipende dagli spettri di assorbimento dei vari cromofori presenti, ma in generale mostra un picco nel rosso e nel blu a causa della presenza ubiquitaria della clorofilla. Il materiale particolato e le sostanze organiche mostrano un comportamento fra loro abbastanza simile, in parte dovuto a similarità di composizione, e in entrambi l'assorbimento decresce in modo approssimativamente esponenziale passando dal blu al rosso. Tirando le somme: per acque oligotrofiche, con concentrazione molto bassa di materia sospesa e dissolta (Caso 1), il coefficiente d'assorbimento è dominato largamente dall'acqua, la lunghezza d'onda di minimo assorbimento è nel blu e l'acqua marina appare di questo colore (cfr. fig. \ref{fgr:graphaw}); per acque eutrofiche, con alta concentrazione di materia sospesa e dissolta (Caso 2), il coefficiente di assorbimento è dominato da questo materiale e il minimo di assorbimento si sposta verso il verde, conferendo questa colorazione a quelle acque.
\subsection{Lo scattering molecolare nei liquidi}
L'assorbimento selettivo, singolarmente considerato, potrebbe forse rendere conto del colore che si osserva quando si è immersi nell’acqua o quando una grande massa d’acqua, come quella che costituisce un ghiacciaio, è interposta tra la sorgente luminosa e l’osservatore, ma non può spiegare perché una massa d’acqua appare di un determinato colore quando la si guarda dall’esterno. È il secondo termine (\texttt{b}) del nostro binomio, la diffusione della radiazione elettromagnetica attraverso il meccanismo di Rayleigh, che permette a una parte della luce - quella trasmessa nell'acqua, in cui predomina la componente blu a causa dell'assorbimento selettivo a grandi lunghezze d'onda - di riemergere dalla superficie facendo sì che vediamo il mare di quel colore. È proprio l'analisi di questo processo che ingenera i maggiori fraintendimenti e meno soddisfa i più esigenti. Per questa ragione, nella nostra discussione dedicheremo alla descrizione dello scattering molecolare uno spazio maggiore di quello che abbiamo riservato all'assorbimento, che è più facile trovare adeguatamente trattato anche altrove. 

I meccanismi fisici fondamentali dello scattering della luce in acqua sono schematizzati nelle figure \ref{fgr:graph03} e \ref{fgr:graph04}. In realtà, il termine "meccanismo di Rayleigh" indica tutta una serie di fenomeni\cite{young1981rayleigh,young1982rayleigh}. Lo scattering (lett. "sparpagliamento") molecolare comprende infatti (v. figura \ref{fgr:graph03}):
\begin{figure}
\includegraphics[width=\textwidth]{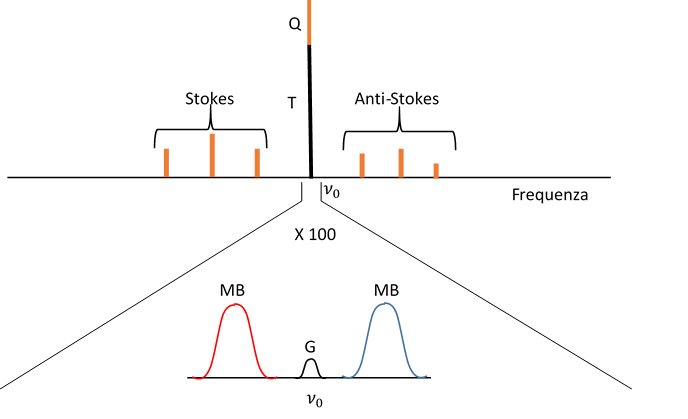}
\caption{La parte coerente T (riga in nero) ha la stessa frequenza $\nu_0$ della radiazione incidente. Gli effetti di rotazione molecolare suddividono la parte incoerente, anisotropa e depolarizzata (in arancio) in una componente non spostata Q e nelle due componenti Stokes e Anti-Stokes (disuguali a causa di effetti quanto-meccanici), che danno origine alle ramificazioni della banda rotazionale Raman. Gli effetti dell'agitazione termica compaiono aumentando la risoluzione della frequenza, per es., di un fattore 100. Il profilo della riga centrale (T+Q) di Cabannes dipende dalla densità: a bassa densità le molecole diffondono la radiazione indipendentemente le une dalle altre, producendo un profilo gaussiano (non rappresentato in figura); ad alta densità la riga centrale di Gross (G, dovuta a scattering da fluttuazioni di densità stazionarie di origine termica) biseca il doppietto di Mandel'shtam--Brillouin (MB, dovuto a scattering da fluttuazioni di densità in movimento, quelle raffigurate come onde sonore nella figura \ref{fgr:graph04}). Le intensità e le larghezze delle tre componenti della riga di Cabannes non sono in scala: le intensità relative dipendono dal rapporto dei calori specifici, le larghezze dalla viscosità e dalla conduttività termica. Poiché l'intensità delle righe generate da questi processi di scattering dipende approssimativamente da $\lambda^{-4}$ il massimo si ha per frequenze della luce incidente nell'estremità blu-violetta dello spettro visibile (Adattato da:~\citeauthor{young1982rayleigh},~\citeyear{young1982rayleigh}).}
\label{fgr:graph03}
\end{figure}
\begin{enumerate}[(a)]
\item L'effetto dell'anisotropia molecolare, che dà origine a due componenti nella luce diffusa: una parte incoerente e depolarizzata (dovuta all'anisotropia) e una parte coerente (dovuta alla parte isotropa della polarizzabilità molecolare). Entrambe le componenti hanno la stessa frequenza della radiazione incidente. La parte coerente costituisce la componente principale della radiazione diffusa.
\item L'effetto della rotazione molecolare, in virtù del quale la componente anisotropa si suddivide in due ramificazioni laterali spostate l'una verso il rosso e l'altra verso il blu (dette, rispettivamente, righe Stokes e anti-Stokes della banda rotazionale Raman), mantenendo però una parte centrale non spostata in lunghezza d'onda (detta riga di Cabannes).
\item L'effetto delle fluttuazioni termiche e meccaniche, che si manifesta nella struttura fine della riga centrale, che si può osservare aumentando fortemente la risoluzione della frequenza. 
\end{enumerate}
Per comprendere questi effetti bisogna considerare che i meccanismi della diffusione della luce nei liquidi sono più complessi di quelli che si manifestano nei gas diluiti. La principale differenza risiede nella natura delle fluttuazioni che originano la diffusione. Queste, nei gas sono di un unico tipo, ovvero le fluttuazioni di densità che intervengono, ad esempio, nella spiegazione del blu del cielo. Nell’acqua, e in altri mezzi condensati, c’è un tipo addizionale di diffusione, dovuto alle cosiddette fluttuazioni di anisotropia, che consistono in aggregazioni temporanee a varie scale di grandezza di gruppi di molecole orientate preferenzialmente in una certa direzione. Infatti, benché l’acqua sia un mezzo isotropo su scala macroscopica, le sue molecole sono otticamente anisotrope e interagendo tra loro possono orientarsi dando origine a questi aggregati (nei gas diluiti le fluttuazioni di anisotropia non possono formarsi perché le molecole sono troppo distanziate per interagire nel modo richiesto). I due tipi di fluttuazione influenzano diversamente la diffusione, per il fatto che la diffusione dovuta alle fluttuazioni di densità polarizza la luce in direzione trasversale rispetto alla direzione di scattering (come nella luce dal cielo) mentre la luce diffusa dalle fluttuazioni di anisotropia è pressoché non polarizzata (le fluttuazioni di anisotropia, infatti, fanno fluttuare localmente la polarizzazione elettrica del mezzo e questo provoca la depolarizzazione della luce diffusa). I due differenti tipi di diffusione spiegano quindi perché, a differenza della luce dal cielo, prevalentemente polarizzata, la luce che proviene dal mare è una mescolanza di luce polarizzata e non polarizzata. Infine, l'effetto delle fluttuazioni causate dai disturbi meccanici si manifesta in un peculiare fenomeno di modulazione della luce nell'interazione con le onde sonore (v. figura \ref{fgr:graph04}).
\begin{figure}
\includegraphics[width=\textwidth]{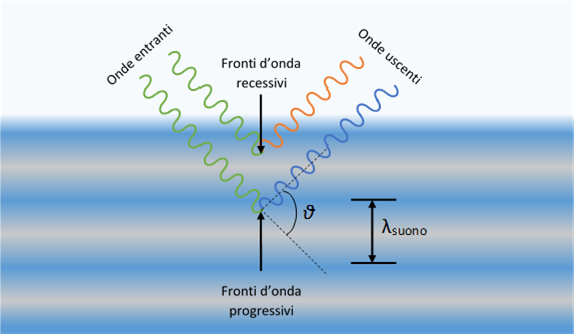}
\caption{I disturbi meccanici creano fluttuazioni di densità che si propagano con la velocità del suono nell'acqua. Possiamo immaginare che la luce incidente di frequenza $\nu_0$ venga spostata per effetto Doppler verso il blu o il rosso, rispettivamente, a causa della diffusione da fronti d'onda approssimativamente in avanzamento o in arretramento. Lo spostamento di frequenza, che determina la comparsa del doppietto MB mostrato nella precedente figura \ref{fgr:graph03}, può essere facilmente calcolato con la legge di Bragg ottenendo $\Delta\nu_{MB}/\nu_0=(v/c)2n\sin\theta/2$ ($v$ velocità del suono nell'acqua, $n$ indice di rifrazione dell'acqua). L'indice di rifrazione dipende da $\nu_0$, dalla salinità e dalla temperatura. (Adattato da:~\citeauthor{dickey2011shedding},~\citeyear{dickey2011shedding}).}
\label{fgr:graph04}
\end{figure}

La prima indagine teorica della modulazione della luce diffusa, realizzata dal fisico russo Mandel'shtam, risale al 1918, sebbene la corrispondente pubblicazione sia apparsa solo nel 1926\cite{mandelstam1926light}. Nel frattempo, il francese Brillouin\cite{brillouin1922diffusion} aveva riscoperto indipendentemente e pubblicato parte dei risultati di Mandel'shtam\cite{fabelinskii2012molecular}. 

Riassumendo, lo scattering molecolare nei liquidi consiste: dello scattering di Rayleigh (i) e dello scattering vibrazionale Raman (ii). A sua volta, lo scattering di Rayleigh produce: le righe rotazionali Raman e la riga centrale di Cabannes. Quest'ultima, infine, è composta: dal doppietto di Mandel'shtam-Brillouin (MB) e dalla riga centrale, detta riga di Gross\cite{gross1930change} (o di Landau-Placzek\cite{landau1934structure}).

Tutti questi effetti coesistono allo stesso tempo e forniscono agli oceanografi importanti informazioni per monitorare (anche da remoto) parametri quali la velocità del suono, la temperatura e la salinità nell'oceano in funzione della profondità. Come lo scattering di Rayleigh nell'atmosfera, entrambi i processi (i) e (ii) dipendono approssimativamente da $\lambda^{-4}$, e questo contribuisce al colore blu del mare, come accade per il cielo blu\footnote{La dipendenza da $\lambda^{-4}$ è solo approssimata perché la teoria di Rayleigh non vale rigorosamente per i liquidi. Sono state proposte alcune espressioni più precise per il coefficiente di scattering $b_w(\lambda)$ in acqua di mare pura ($b_w$ in \SI{}{\per\meter}, $\lambda$ in \si{\nano\meter}), la più utilizzata delle quali è la seguente\cite{CHEN201843}:
\begin{equation}
   b_w(\lambda)=0.00288(\frac{\lambda}{500})^{-4.32}
   \label{eqn:footed} \nonumber
\end{equation}}.
La dipendenza approssimata da $\lambda^{-4}$ si può giustificare in modo intuitivo con semplici considerazioni di tipo dimensionale (cfr. rif.~\citenum{adam2011mathematical}). Il ragionamento si può riassumere in questo modo. L'intensità della radiazione diffusa è proporzionale all'intensità della radiazione incidente attraverso un fattore di proporzionalità (adimensionale) che dipende dal volume $V$ della particella diffondente, dalla distanza $r$ dal centro diffusore al punto di osservazione, dalla lunghezza d'onda $\lambda$ della radiazione diffusa, dagli indici di rifrazione all'interno e all'esterno del diffusore. Essendo per definizione adimensionali gli indici di rifrazione escono dal ragionamento. $V$ ha le dimensioni del cubo di una lunghezza, $r$ e $\lambda$ della prima potenza di una lunghezza. Poiché l'ampiezza della radiazione diffusa è proporzionale al numero dei centri diffusori, a sua volta proporzionale al loro volume aggregato, l'intensità, cioè il quadrato dell'ampiezza, deve avere le dimensioni del quadrato di un volume ovvero della sesta potenza di una lunghezza $([L]^6)$. Inoltre, poiché per un dipolo (una molecola con le cariche positive e negative non coincidenti che costituiscono un tipico centro diffusore), a grande distanza, vale la legge dell'inverso del quadrato per l'energia irradiata per unità di area, l'intensità diffusa dipende da $r^{-2}$, ovvero l'inverso della seconda potenza di una lunghezza $([L]^{-2})$. Esprimendo la relazione tra gli esponenti come $6+(-2)+\alpha=0$, dove $\alpha$ è l'esponente di $\lambda$, otteniamo $\alpha=-4$ quindi $I_{scatt}\propto\lambda^{-4}$.

Un altro fatto facilmente giustificabile -- ma spesso sottaciuto -- è che in assenza di inomogeneità il mare apparirebbe nero se osservato da qualunque direzione eccetto quella di propagazione della luce trasmessa direttamente (senza diffusione). In altre parole, la diffusione non si verifica in un mezzo perfettamente omogeneo, ossia privo di impurità o di fluttuazioni della densità molecolare. Una giustificazione elementare di ciò per un gas diluito si trova ad es. in rif.~\citenum{aloisienigma} applicata al cielo, e più in generale per i mezzi omogenei (liquidi e gas) in rif.~\citenum{fabelinskii2012molecular}, pp.~5--6.

Vogliamo aggiungere un'ultima annotazione sulle caratteristiche dei fenomeni di diffusione della luce in mare analoghi a quelli in atmosfera che dipendono dalle dimensioni dei centri diffusori:
\begin{enumerate}[(a)]
\item al primo ordine, cioè per dimensioni dell’ordine della lunghezza d’onda della luce incidente (molecole d’acqua e sali disciolti), prevale la diffusione di Rayleigh, caratterizzata da distribuzione angolare simmetrica avanti-indietro (cioè angolo medio di diffusione nullo) e piccola intensità diffusa con un angolo piccolo rispetto alla direzione incidente;
\item al secondo ordine, per dimensioni superiori alla lunghezza d’onda incidente (materiale particolato organico e inorganico) diventa importante la riflessione diffusa o diffusione di Mie, asimmetrica, con picco pronunciato in avanti e caratterizzata da piccoli angoli di deflessione;
\item il caso delle bolle d'aria, che possono raggiungere dimensioni macroscopiche, viene trattato con la teoria di Mie o, per particelle di dimensioni ancora maggiori, con l'approssimazione dell'ottica geometrica. 
\end{enumerate}
Nel seguito useremo preferenzialmente il termine \textit{scattering} per riferirci alla diffusione al primo ordine (molecolare) e il termine \textit{diffusione} per le altre particelle.
\section{Conclusioni: un "mare modello" di test}
Tutta l'analisi finora esposta può essere codificata in un modello fenomenologico piuttosto semplice, che consente di mettere in relazione le osservazioni fondamentali con la teoria. Lo strumento rappresentato nella figura \ref{fgr:graph02} mostra all'osservatore, attraverso l'oculare \textit{O}, due spettri sovrapposti: lo spettro della luce che penetra nel mare direttamente dall'alto (dal cielo e dal sole), e quello della luce che risale dalle profondità marine. L'apparato è schermato dal riflesso della luce dal cielo in modo da eliminarne l'influenza. 

Il modello considera la riflessione diffusa uniforme dalla superficie marina, considerata come una superficie ideale perfettamente diffondente (superficie lambertiana), che obbedisce alla cosiddetta legge del pi greco o legge di Lambert. Per superficie di questo tipo la luminanza $L$ (intensità per unità di superficie emettente) emessa o riflessa è uguale per ogni direzione. L'emettenza $E$ (flusso luminoso per unità di superficie irradiata) è legata alla luminanza dalla legge del pi greco: $L=E/\pi$

Il rapporto tra l'intensità $L$ della radiazione emessa perpendicolarmente alla superficie del mare e l'intensità incidente $I_0$ si può esprimere con la formula seguente, che non dimostriamo (per una dimostrazione cfr. rif.~\mciteSubRef{shoulejkin1923color}) ma di cui esponiamo una discussione in termini qualitativi:
\begin{equation}
  \frac{L}{I_0}=\frac{1}{\pi}\frac{(1-\beta)\cdot\frac{1}{2} b_w(\lambda)+\beta b_p(\lambda)}{(1-\beta) \cdot \frac{1}{2}b_w(\lambda)+2a_w(\lambda)+\beta[2-b_p(\lambda)]} \label{eqn:general} 
\end{equation}

Nell'equazione \ref{eqn:general} compare al numeratore e al denominatore il coefficiente di scattering (molecolare) $b_w$ per l'acqua di mare pura che segue approssimativamente la legge di Rayleigh. Poiché, se denotiamo con $\theta$ l'angolo di scattering, l'energia del raggio diffuso è proporzionale a $(1+\cos^2{\theta})$, quindi la stessa per $\theta$ e $-\theta$, tutti i centri diffusori contenuti in uno strato d'acqua alla stessa profondità diffonderanno un'eguale quantità di energia sia al di sopra che al di sotto di questo strato: da ciò deriva la frazione $\nicefrac{1}{2}$ che moltiplica il termine di Rayleigh e che rappresenta il contributo di backscattering. Il coefficiente di assorbimento $a_w(\lambda)$ dipende dalla lunghezza d'onda (assorbimento selettivo), come pure $b_p(\lambda)$, che rappresenta il coefficiente di diffusione (selettiva) per le particelle del secondo ordine. Questo tipo di particelle avrà, corrispondentemente, un coefficiente di assorbimento pari a $1-b_p(\lambda)$, assumendo che tutta la luce intercettata da una particella sia dispersa o assorbita. Infine, $\beta$ rappresenta la probabilità che un raggio incontri una particella del secondo ordine in un percorso unitario (e $1-\beta$ che non la incontri, e in questo caso il contributo all'energia diffusa sarà dovuto al primo ordine che sarà quindi proporzionale a questa probabilità). Si vede quindi che al numeratore compare l'energia risalente dalla superficie del mare (e raccolta dallo strumento), dovuta a (back)scattering dal primo ordine (molecole, dimensioni $\approx\lambda$) e a diffusione o riflessione dal secondo ordine (particelle, bolle, dimensioni $\gg\lambda$), con pesi, rispettivamente $1-\beta$ e $\beta$. Al denominatore abbiamo invece l'energia che attraversa lo strumento dall'alto, comprendente tutta l'energia che dopo aver interagito risalirà dalla superficie verso lo strumento (la parte che ricompare al numeratore) più quella assorbita nell'interazione dalla massa d'acqua e dalle particelle disperse.

Osserviamo che in acque molto pure, ossia in assenza di particelle sospese $(\beta=0)$, la \ref{eqn:general} si riduce a: 
\begin{equation}
  \frac{L}{I_0}=\frac{1}{\pi}\frac{\frac{1}{4} b_w(\lambda)}{\frac{1}{4}b_w(\lambda)+a_w(\lambda)} \label{eqn:reduced} 
\end{equation}
Da questa equazione si vede che sia la diffusione che l'assorbimento sono necessari per spiegare il colore del mare. Infatti, in assenza di diffusione $(b_w(\lambda)=0)$ si otterrebbe una superficie marina perfettamente nera $(L=0)$, mentre senza assorbimento $(a_w(\lambda)=0)$ la superficie sarebbe perfettamente bianca perché $L=I_0/\pi=costante$, indipendente da $\lambda$.

\begin{figure}
\includegraphics[width=\textwidth]{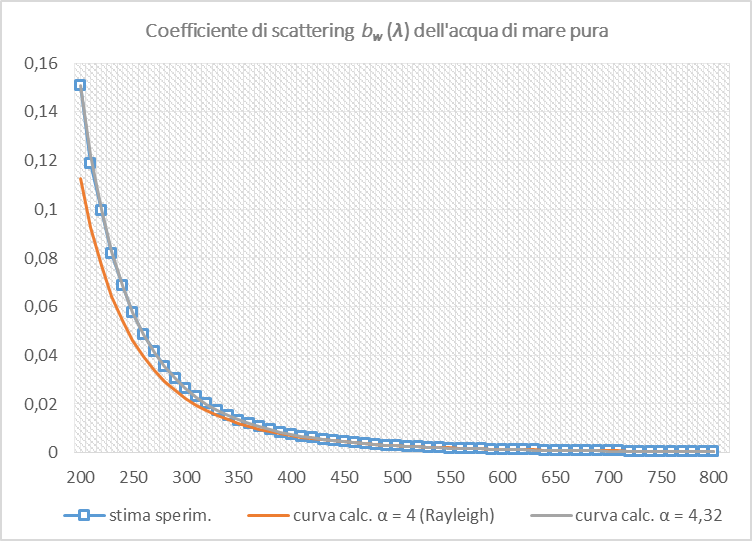}
\caption{Le curve del coefficiente di scattering $b_w(\lambda)$ dell'acqua di mare pura sono calcolate utilizzando la legge di Rayleigh ${b_w=b_0(\lambda/\lambda_0)^{-\alpha}}$ con $\alpha=4$ (curva in rosso) e con $\alpha=4,32$ (curva in grigio); ${\lambda_0=\SI{500}{\nano\meter}}$, $b(0)=\SI{0,00288}{\per\meter}$. La seconda curva è pressoché indistinguibile dalla migliore stima sperimentale riportata in \citeauthor{smith1981optical}, \citeyear{smith1981optical}.}
\label{fgr:graphbw}
\end{figure}

Shoulejkin (cfr. rif.~\mciteSubRef{shoulejkin1923color}) testò questo modello assumendo la validità della legge di Rayleigh ${b_w=b_0(\lambda/\lambda_0)^{-\alpha}}$ nella sua forma originaria (valida per i gas) con $\alpha=4$. Fissando una lunghezza d'onda di riferimento (ad es. ${\lambda_0=\SI{500}{\nano\meter}}$) e il valore del parametro $b_0$ misurato sperimentalmente (ad es. $b(\SI{500}{\nano\meter})=\SI{0,00288}{\per\meter}$) e poi cambiando questi valori, il modello permette di tracciare vari tipi di curve (di attenuazione e spettrali), che risultano in discreto accordo qualitativo con le migliori misure disponibili all'epoca al ricercatore sovietico nel range (\SI{480}{\nano\meter}-\SI{660}{\nano\meter}); misure, queste, che erano state ottenute dal Conte Aufsess a inizio secolo\cite{von2013physikalischen,*frhr1904farbe}. Se invece si utilizza la legge di Rayleigh nella versione moderna per l'acqua pura di mare, con la pendenza della curva di scattering modificata ($\alpha=4,32$), si trova che l'accordo con i dati sperimentali è pressoché completo (v. fig.\ref{fgr:graphbw}). 
Il coefficiente di attenuazione diffusa (assorbimento + backscattering), definito come $K_w=a_w+\frac{1}{2}b_w$, è tuttavia così piccolo che risultava (e risulta tuttora) estremamente difficoltoso fare gli esperimenti in laboratorio con acqua naturale (o anche con acqua marina artificiale appositamente preparata). Shoulejkin superò questo problema usando un liquido più intensamente colorato dell'acqua e aumentando la concentrazione delle particelle sospese. Dopo lunghi e tediosi tentativi trovò che una sostanza colorante accettabile per gli scopi del test, in virtù del suo spettro di assorbimento e la rassomiglianza con l'acqua, era il bleu rodolina (Rhodulin blue) e la sospensione più adatta la soluzione colloidale di colofonia (o rosina) in alcol, quando ne venivano aggiunte piccole gocce all'acqua colorata dalla rodolina. Ancora una volta, la concordanza tra le curve calcolate e i risultati sperimentali si dimostrò abbastanza buona per la soluzione di rodolina, e in accordo qualitativo per la stessa soluzione con le particelle di rosina in sospensione\footnote{In realtà l'accordo con i dati sperimentali non è così buono se si estrapolano questi calcoli alle lunghezze d'onda inferiori a circa \SI{460}{\nano\meter}, non esplorate da Shoulejkin, perché alle piccole lunghezze d'onda diventano importanti le deviazioni dalle legge di Rayleigh. Non è chiaro se all'epoca il ricercatore sovietico fosse consapevole di questo problema, resta il fatto che nella sua memoria originale del 1923 (cfr. rif~\mciteSubRef{shoulejkin1923color}) il confronto con le curve sperimentali è limitato alla regione dello spettro visibile al di sopra dei \SI{460}{\nano\meter}, dove erano in effetti disponibili dati attendibili e dove la legge di Rayleigh funziona abbastanza bene.}.
\begin{figure}
\includegraphics[width=\textwidth]{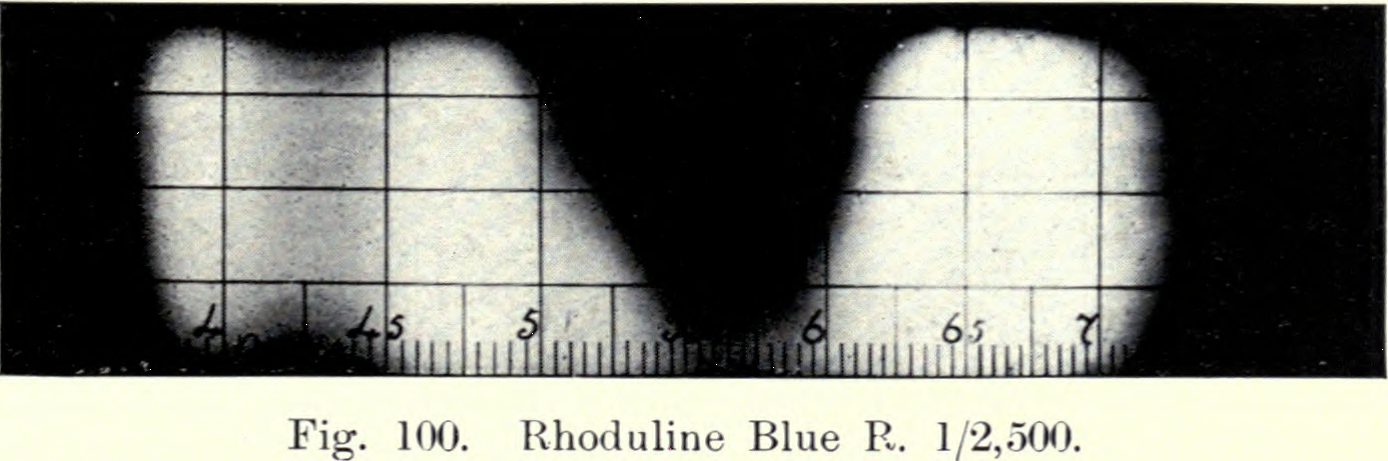}
\caption{Spettro di assorbimento del bleu rodolina\cite{mees1909atlas}.}
\label{fgr:RhodSpectra}
\end{figure}
Si potrebbe pensare che al giorno d'oggi la dipendenza dell'attenuazione dalla lunghezza d'onda sia nota con alta precisione. In realtà, se ciò può essere vero per la parte dovuta allo scattering, le misure del coefficiente di assorbimento ottenute negli esperimenti moderni, benché molto migliorate grazie all'utilizzo di tecniche molto sofisticate, evidenziano ancora notevoli discrepanze vicino al minimo di assorbimento a \SI{420}{\nano\meter} nel blu dello spettro visibile, dove la luce può percorrere altre cento metri con piccolo assorbimento, rendendo difficili le misure di laboratorio. L'assorbimento cresce di quasi mille volte passando dal minimo nel blu al massimo nel rosso. L'assorbimento molto piccolo sulla scala dei metri nel blu è la causa delle discrepanze così grandi delle misure alla fine dello spettro visibile\bibnote{Approfondimenti sulla questione si possono trovare al link: \url{http://www1.lsbu.ac.uk/water/water_sitemap.html}; un'ottima risorsa è anche l'\texttt{Optical Absorption of Water Compendium} al link: \url{https://omlc.org/spectra/water/abs/index.html}}.

\begin{figure}
\includegraphics[width=\textwidth]{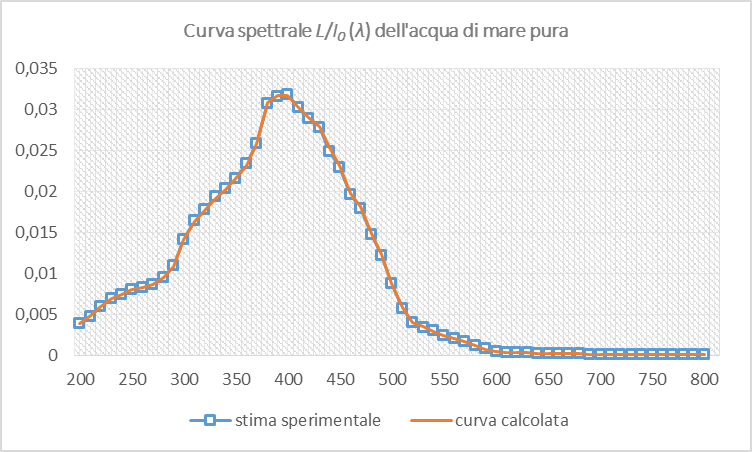}
\caption{Curva spettrale in funzione di $\lambda$ (in \SI{}{\nano\meter}) calcolata mediante l'equazione \ref{eqn:reduced}, trascurando l'effetto delle particelle sospese. Come dati di partenza vengono utilizzati i valori del coefficiente di assorbimento della curva di figura \ref{fgr:graphaw} e del coefficiente di scattering della curva di figura \ref{fgr:graphbw} calcolati per $\alpha=4,32$. Il massimo della curva spettrale si ha in corrispondenza di $\lambda=\SI{400}{\nano\meter}$. Incidentalmente, la curva mostra anche che nella regione del visibile l'energia riflessa dal mare verso lo spazio è una piccola frazione dell'energia incidente.}
\label{fgr:graphspectra}
\end{figure}

Non si può, ovviamente, pretendere di riprodurre questo genere di esperimenti in un laboratorio scolastico. È invece molto semplice utilizzare in aula programmi come \texttt{Microsoft Excel™} per verificare la legge di Rayleigh nei liquidi e mostrare gli andamenti di varie curve sperimentali (v. figg. \ref{fgr:graphaw}, \ref{fgr:graphbw}, \ref{fgr:graphspectra}). L'analisi di queste curve permette, meglio di altre spiegazioni, di illustrare e verificare (almeno nel caso più semplice delle acque di Caso 1) tutti i principali concetti esposti in questo articolo.
\bibliography{achemso-paper}

\providecommand{\latin}[1]{#1}
\makeatletter
\providecommand{\doi}
  {\begingroup\let\do\@makeother\dospecials
  \catcode`\{=1 \catcode`\}=2\doi@aux}
\providecommand{\doi@aux}[1]{\endgroup\texttt{#1}}
\makeatother
\providecommand*\mcitethebibliography{\thebibliography}
\csname @ifundefined\endcsname{endmcitethebibliography}
  {\let\endmcitethebibliography\endthebibliography}{}
\begin{mcitethebibliography}{76}
\providecommand*\natexlab[1]{#1}
\providecommand*\mciteSetBstSublistMode[1]{}
\providecommand*\mciteSetBstMaxWidthForm[2]{}
\providecommand*\mciteBstWouldAddEndPuncttrue
  {\def\EndOfBibitem{\unskip.}}
\providecommand*\mciteBstWouldAddEndPunctfalse
  {\let\EndOfBibitem\relax}
\providecommand*\mciteSetBstMidEndSepPunct[3]{}
\providecommand*\mciteSetBstSublistLabelBeginEnd[3]{}
\providecommand*\EndOfBibitem{}
\mciteSetBstSublistMode{f}
\mciteSetBstMaxWidthForm{subitem}{(\alph{mcitesubitemcount})}
\mciteSetBstSublistLabelBeginEnd
  {\mcitemaxwidthsubitemform\space}
  {\relax}
  {\relax}

\bibitem[Aloisi and Nali(2006)Aloisi, and Nali]{aloisienigma}
Aloisi,~A.~M.; Nali,~P.~F. L'enigma del cielo azzurro. 2006; disponibile
  all'url:
  \url{https://www.researchgate.net/publication/281455615_L'enigma_del_cielo_azzurro}\relax
\mciteBstWouldAddEndPuncttrue
\mciteSetBstMidEndSepPunct{\mcitedefaultmidpunct}
{\mcitedefaultendpunct}{\mcitedefaultseppunct}\relax
\EndOfBibitem
\bibitem[Costanzo(20 aprile 2007)]{costanzo}
Costanzo,~A. Com. priv., 20 aprile 2007\relax
\mciteBstWouldAddEndPuncttrue
\mciteSetBstMidEndSepPunct{\mcitedefaultmidpunct}
{\mcitedefaultendpunct}{\mcitedefaultseppunct}\relax
\EndOfBibitem
\bibitem[v.~ad~es. Scarpa(2004)]{scarpa2004arcobaleno}
v.~ad~es. Scarpa,~F., Ed. \emph{Un arcobaleno di domande. 99 risposte per
  conoscere la scienza}; Nuova biblioteca Dedalo; Dedalo, 2004; p 128\relax
\mciteBstWouldAddEndPuncttrue
\mciteSetBstMidEndSepPunct{\mcitedefaultmidpunct}
{\mcitedefaultendpunct}{\mcitedefaultseppunct}\relax
\EndOfBibitem
\bibitem[Una sintesi molto accessibile si trova in questo bel~libro:
  Bohren(2001)]{bohren2001clouds}
Una sintesi molto accessibile si trova in questo bel~libro: Bohren,~C.~F.
  \emph{Clouds in a Glass of Beer: Simple Experiments in Atmospheric Physics};
  (Wiley science editions); Dover Publications, 2001\relax
\mciteBstWouldAddEndPuncttrue
\mciteSetBstMidEndSepPunct{\mcitedefaultmidpunct}
{\mcitedefaultendpunct}{\mcitedefaultseppunct}\relax
\EndOfBibitem
\bibitem[oce()]{oceanopticsbook}
Tra le migliori risorse online citiamo l'Ocean Optics Web Book,
  \url{http://www.oceanopticsbook.info/}. Originariamente pensato per una
  comunità di specialisti, è una risorsa pubblicamente accessibile che
  risponde alle esigenze educative e di consultazione di un'ampia platea di
  studiosi o di semplici appassionati.\relax
\mciteBstWouldAddEndPunctfalse
\mciteSetBstMidEndSepPunct{\mcitedefaultmidpunct}
{}{\mcitedefaultseppunct}\relax
\EndOfBibitem
\bibitem[v.~ad~es. Monk(2004)]{monk2004physical}
v.~ad~es. Monk,~P. M.~S. \emph{Physical Chemistry: Understanding our Chemical
  World}; Wiley, 2004; a pag. 483 il colore blu del Mediterraneo viene
  attribuito all'effetto Raman. In realtà tale effetto è troppo debole
  perché il problema del colore del mare sia interpretabile in questi
  termini\relax
\mciteBstWouldAddEndPuncttrue
\mciteSetBstMidEndSepPunct{\mcitedefaultmidpunct}
{\mcitedefaultendpunct}{\mcitedefaultseppunct}\relax
\EndOfBibitem
\bibitem[Preisendorfer(1976)]{preisendorfer1976hydrologic}
Preisendorfer,~R. \emph{Hydrologic optics (6 vols)}; U.S. Dept. of Commerce,
  National Oceanic and Atmospheric Administration, Environmental Research
  Laboratories, Pacific Marine Environmental Laboratory, 1976; fuori commercio,
  è un'estesa trattazione di oltre 1700 pagine suddivise in sei volumi,
  matematicamente rigorosa anche se un po' datata\relax
\mciteBstWouldAddEndPuncttrue
\mciteSetBstMidEndSepPunct{\mcitedefaultmidpunct}
{\mcitedefaultendpunct}{\mcitedefaultseppunct}\relax
\EndOfBibitem
\bibitem[Mobley(1994)]{mobley1994light}
Mobley,~C.~D. \emph{Light and water: radiative transfer in natural waters (San
  Diego, CA: Academic)}; Academic press, 1994; fuori commercio, disponibile in
  edizione CD, è un compendio del precedente\relax
\mciteBstWouldAddEndPuncttrue
\mciteSetBstMidEndSepPunct{\mcitedefaultmidpunct}
{\mcitedefaultendpunct}{\mcitedefaultseppunct}\relax
\EndOfBibitem
\bibitem[Per un resoconto storico dal 1600 al 1930~v.
  Wernand(2013)]{WERNAND201335}
Per un resoconto storico dal 1600 al 1930~v. Wernand,~M.~R. In \emph{Subsea
  Optics and Imaging}; Watson,~J., Zielinski,~O., Eds.; Woodhead Publishing
  Series in Electronic and Optical Materials; Woodhead Publishing, 2013; pp
  35--79\relax
\mciteBstWouldAddEndPuncttrue
\mciteSetBstMidEndSepPunct{\mcitedefaultmidpunct}
{\mcitedefaultendpunct}{\mcitedefaultseppunct}\relax
\EndOfBibitem
\bibitem[cla()]{clarityonthesea}
segnaliamo inoltre il sito \url{http://clarityonthesea.org}, che offre
  un'interessante raccolta di pubblicazioni storiche sul colore e la limpidezza
  del mare\relax
\mciteBstWouldAddEndPuncttrue
\mciteSetBstMidEndSepPunct{\mcitedefaultmidpunct}
{\mcitedefaultendpunct}{\mcitedefaultseppunct}\relax
\EndOfBibitem
\bibitem[Bunsen(1847)]{bunsen1847ueber}
Bunsen,~R. \emph{Justus Liebigs Ann. Chem.} \textbf{1847}, \emph{62},
  1--59\relax
\mciteBstWouldAddEndPuncttrue
\mciteSetBstMidEndSepPunct{\mcitedefaultmidpunct}
{\mcitedefaultendpunct}{\mcitedefaultseppunct}\relax
\EndOfBibitem
\bibitem[Bunsen(1847)]{bunsen1847blaue}
Bunsen,~R. \emph{Jahresber. Fortschr. Chem.} \textbf{1847}, \emph{1847},
  1236\relax
\mciteBstWouldAddEndPuncttrue
\mciteSetBstMidEndSepPunct{\mcitedefaultmidpunct}
{\mcitedefaultendpunct}{\mcitedefaultseppunct}\relax
\EndOfBibitem
\bibitem[Bunsen(1849)]{bunsen1849colour}
Bunsen,~R. \emph{Edinburgh New Philos. J.} \textbf{1849}, \emph{47},
  95--98\relax
\mciteBstWouldAddEndPuncttrue
\mciteSetBstMidEndSepPunct{\mcitedefaultmidpunct}
{\mcitedefaultendpunct}{\mcitedefaultseppunct}\relax
\EndOfBibitem
\bibitem[Bancroft(1919)]{bancroft1919color}
Bancroft,~W.~D. \emph{J. Franklin Inst.} \textbf{1919}, \emph{187}, 459--485,
  che contiene anche una rassegna di articoli sul colore del mare pubblicati
  prima del 1917\relax
\mciteBstWouldAddEndPuncttrue
\mciteSetBstMidEndSepPunct{\mcitedefaultmidpunct}
{\mcitedefaultendpunct}{\mcitedefaultseppunct}\relax
\EndOfBibitem
\bibitem[Tyndall(1869)]{tyndall1869iv}
Tyndall,~J. \emph{Proc. R. Soc. Lond.} \textbf{1869}, \emph{17}, 223--233\relax
\mciteBstWouldAddEndPuncttrue
\mciteSetBstMidEndSepPunct{\mcitedefaultmidpunct}
{\mcitedefaultendpunct}{\mcitedefaultseppunct}\relax
\EndOfBibitem
\bibitem[Tyndall(1870)]{tyndall1870colour}
Tyndall,~J. \emph{Nature} \textbf{1870}, \emph{2}, 489--490\relax
\mciteBstWouldAddEndPuncttrue
\mciteSetBstMidEndSepPunct{\mcitedefaultmidpunct}
{\mcitedefaultendpunct}{\mcitedefaultseppunct}\relax
\EndOfBibitem
\bibitem[Tyndall(1871)]{tyndall1871colour}
Tyndall,~J. On the colour of water and the scattering of light in water and
  air. Proc. R. Inst. 1871; pp 188--199\relax
\mciteBstWouldAddEndPuncttrue
\mciteSetBstMidEndSepPunct{\mcitedefaultmidpunct}
{\mcitedefaultendpunct}{\mcitedefaultseppunct}\relax
\EndOfBibitem
\bibitem[M'Master(1871)]{tyndall1871causes}
M'Master,~W. \emph{Nature} \textbf{1871}, \emph{4}, 203--204\relax
\mciteBstWouldAddEndPuncttrue
\mciteSetBstMidEndSepPunct{\mcitedefaultmidpunct}
{\mcitedefaultendpunct}{\mcitedefaultseppunct}\relax
\EndOfBibitem
\bibitem[Tyndall(1886)]{tyndall1886fragments}
Tyndall,~J. \emph{Fragments of science: A series of detached essays, addresses
  and reviews}; Appleton, 1886; Vol.~1; pp 142--174\relax
\mciteBstWouldAddEndPuncttrue
\mciteSetBstMidEndSepPunct{\mcitedefaultmidpunct}
{\mcitedefaultendpunct}{\mcitedefaultseppunct}\relax
\EndOfBibitem
\bibitem[v.~ad~es. Tyndall(6 aprile 1871)]{tyndall_1871Sydney}
v.~ad~es. Tyndall,~J. Professor Tyndall on the London Water Supply. (From the
  Times.) \textit{Sidney Morning Herald}. 6 aprile 1871; p 6\relax
\mciteBstWouldAddEndPuncttrue
\mciteSetBstMidEndSepPunct{\mcitedefaultmidpunct}
{\mcitedefaultendpunct}{\mcitedefaultseppunct}\relax
\EndOfBibitem
\bibitem[Tyndall(27 gennaio 1871)]{tyndall_1871Engineer}
Tyndall,~J. (Annotato da un redattore dell'Engineer durante una lezione).
  Professor Tyndall on the colour of the sea and the water supply of London.
  \textit{The Engineer}. 27 gennaio 1871; p 64, cit.
  in~\mciteSubRef{WERNAND201335}\relax
\mciteBstWouldAddEndPuncttrue
\mciteSetBstMidEndSepPunct{\mcitedefaultmidpunct}
{\mcitedefaultendpunct}{\mcitedefaultseppunct}\relax
\EndOfBibitem
\bibitem[Tyndall(31 gennaio 1871)]{tyndall_1871Times}
Tyndall,~J. Professor Tyndall on the London Water Supply. \textit{The Times}.
  31 gennaio 1871; p 4, cit. in~\citenum{jackson2018ascent}\relax
\mciteBstWouldAddEndPuncttrue
\mciteSetBstMidEndSepPunct{\mcitedefaultmidpunct}
{\mcitedefaultendpunct}{\mcitedefaultseppunct}\relax
\EndOfBibitem
\bibitem[Jackson(2018)]{jackson2018ascent}
Jackson,~R. \emph{The Ascent of John Tyndall: Victorian Scientist, Mountaineer,
  and Public Intellectual}; OUP Oxford, 2018\relax
\mciteBstWouldAddEndPuncttrue
\mciteSetBstMidEndSepPunct{\mcitedefaultmidpunct}
{\mcitedefaultendpunct}{\mcitedefaultseppunct}\relax
\EndOfBibitem
\bibitem[O'Connell(3 agosto 2000)]{o'connell_2000}
O'Connell,~S. Britain's leading light. \textit{The Guardian}. 3 agosto 2000;
  \url{https://www.theguardian.com/science/2000/aug/03/technology1}\relax
\mciteBstWouldAddEndPuncttrue
\mciteSetBstMidEndSepPunct{\mcitedefaultmidpunct}
{\mcitedefaultendpunct}{\mcitedefaultseppunct}\relax
\EndOfBibitem
\bibitem[Aitken(1882)]{aitken18821}
Aitken,~J. \emph{Proc. R. Soc. Edinb.} \textbf{1882}, \emph{11}, 472--483\relax
\mciteBstWouldAddEndPuncttrue
\mciteSetBstMidEndSepPunct{\mcitedefaultmidpunct}
{\mcitedefaultendpunct}{\mcitedefaultseppunct}\relax
\EndOfBibitem
\bibitem[Aitken(1899)]{aitken1899colour}
Aitken,~J. \emph{Nature} \textbf{1899}, \emph{59}, 509\relax
\mciteBstWouldAddEndPuncttrue
\mciteSetBstMidEndSepPunct{\mcitedefaultmidpunct}
{\mcitedefaultendpunct}{\mcitedefaultseppunct}\relax
\EndOfBibitem
\bibitem[Not()]{Note-1}
Spring dedicò una serie di pubblicazioni nell'arco di vari anni al colore
  delle acque marine e lacustri. La maggior parte delle opere di Spring sono
  accessibili online sul repository dell’università di Liegi, all'url
  \url{https://orbi.uliege.be/browse?type=authorulg&rpp=20&value=Spring\%2C+Walth\%C3\%A8re+p00022}.
  Un elenco di un centinaio di pubblicazioni si trova all'url
  \url{http://waltherespring1848-1911.e-monsite.com/pages/biographie-de-walthere-spring.html}.
  Le sue \textit{Oeuvres complètes} sono state pubblicate dalla Société
  Chimique de Belgique in 2 volumi (Bruxelles, 1914–1923) con una prefazione
  biografica tratta da Crismer L. \textit{Walthère Spring: sa vie et son
  œuvre}; Ad. Hoste, 1912.\relax
\mciteBstWouldAddEndPunctfalse
\mciteSetBstMidEndSepPunct{\mcitedefaultmidpunct}
{}{\mcitedefaultseppunct}\relax
\EndOfBibitem
\bibitem[Lionetti and Mager(1951)Lionetti, and Mager]{lionetti1951walter}
Lionetti,~F.; Mager,~M. \emph{J. Chem. Educ.} \textbf{1951}, \emph{28},
  604\relax
\mciteBstWouldAddEndPuncttrue
\mciteSetBstMidEndSepPunct{\mcitedefaultmidpunct}
{\mcitedefaultendpunct}{\mcitedefaultseppunct}\relax
\EndOfBibitem
\bibitem[Spring(1883)]{spring1883popular}
Spring,~W. \emph{The Popular Science Monthly} \textbf{1883}, \emph{23},
  68--74\relax
\mciteBstWouldAddEndPuncttrue
\mciteSetBstMidEndSepPunct{\mcitedefaultmidpunct}
{\mcitedefaultendpunct}{\mcitedefaultseppunct}\relax
\EndOfBibitem
\bibitem[De~Thierry(1886)]{DeThierry1886LaNature}
De~Thierry,~M. \emph{La Nature. Revue des sciences} \textbf{1886}, \emph{28},
  3--6\relax
\mciteBstWouldAddEndPuncttrue
\mciteSetBstMidEndSepPunct{\mcitedefaultmidpunct}
{\mcitedefaultendpunct}{\mcitedefaultseppunct}\relax
\EndOfBibitem
\bibitem[Battelli and Pandolfi(1899)Battelli, and Pandolfi]{battelli1899sull}
Battelli,~A.; Pandolfi,~M. \emph{Il Nuovo Cimento (1895-1900)} \textbf{1899},
  \emph{9}, 321--326\relax
\mciteBstWouldAddEndPuncttrue
\mciteSetBstMidEndSepPunct{\mcitedefaultmidpunct}
{\mcitedefaultendpunct}{\mcitedefaultseppunct}\relax
\EndOfBibitem
\bibitem[Demar{\'e}e \latin{et~al.}(2009)Demar{\'e}e, Brouyaux, and
  Verheyden]{demaree2009walthere}
Demar{\'e}e,~G.; Brouyaux,~F.; Verheyden,~R. \emph{Ciel et Terre}
  \textbf{2009}, \emph{125}\relax
\mciteBstWouldAddEndPuncttrue
\mciteSetBstMidEndSepPunct{\mcitedefaultmidpunct}
{\mcitedefaultendpunct}{\mcitedefaultseppunct}\relax
\EndOfBibitem
\bibitem[Gillispie(1981)]{gillispie1970dictionary}
Gillispie,~C. C. e. i.~c. \emph{Dictionary of Scientific Biography. Volume 12.
  IBN RUSHD - JEAN-SERVAIS STAS}; Dictionary of Scientific Biography; Charles
  Scribner's Sons, 1981; Vol.~12; pp 592--594\relax
\mciteBstWouldAddEndPuncttrue
\mciteSetBstMidEndSepPunct{\mcitedefaultmidpunct}
{\mcitedefaultendpunct}{\mcitedefaultseppunct}\relax
\EndOfBibitem
\bibitem[Plass \latin{et~al.}(1978)Plass, Humphreys, and
  Kattawar]{plass1978color}
Plass,~G.~N.; Humphreys,~T.~J.; Kattawar,~G.~W. \emph{Appl. Opt.}
  \textbf{1978}, \emph{17}, 1432--1446\relax
\mciteBstWouldAddEndPuncttrue
\mciteSetBstMidEndSepPunct{\mcitedefaultmidpunct}
{\mcitedefaultendpunct}{\mcitedefaultseppunct}\relax
\EndOfBibitem
\bibitem[Ricc{\`o}(1876)]{ricco1876studi}
Ricc{\`o},~A. \emph{Mem. Soc. Spett. Ital.} \textbf{1876}, \emph{5},
  A101--A115\relax
\mciteBstWouldAddEndPuncttrue
\mciteSetBstMidEndSepPunct{\mcitedefaultmidpunct}
{\mcitedefaultendpunct}{\mcitedefaultseppunct}\relax
\EndOfBibitem
\bibitem[Ricc{\`o}(1879)]{ricco1879studi}
Ricc{\`o},~A. \emph{Mem. Soc. Spett. Ital.} \textbf{1879}, \emph{8},
  A1--A10\relax
\mciteBstWouldAddEndPuncttrue
\mciteSetBstMidEndSepPunct{\mcitedefaultmidpunct}
{\mcitedefaultendpunct}{\mcitedefaultseppunct}\relax
\EndOfBibitem
\bibitem[Ricc{\`o}(1904)]{ricco1904sul}
Ricc{\`o},~A. \emph{Mem. Soc. Spett. Ital.} \textbf{1904}, \emph{33},
  106--111\relax
\mciteBstWouldAddEndPuncttrue
\mciteSetBstMidEndSepPunct{\mcitedefaultmidpunct}
{\mcitedefaultendpunct}{\mcitedefaultseppunct}\relax
\EndOfBibitem
\bibitem[Strutt(1910)]{rayleigh1910colours}
Strutt,~J.~W. \emph{Nature} \textbf{1910}, \emph{83}, 48--50\relax
\mciteBstWouldAddEndPuncttrue
\mciteSetBstMidEndSepPunct{\mcitedefaultmidpunct}
{\mcitedefaultendpunct}{\mcitedefaultseppunct}\relax
\EndOfBibitem
\bibitem[Raman(1922)]{raman1922molecular}
Raman,~C.~V. \emph{Proc. R. Soc. Lond. A} \textbf{1922}, \emph{101},
  64--80\relax
\mciteBstWouldAddEndPuncttrue
\mciteSetBstMidEndSepPunct{\mcitedefaultmidpunct}
{\mcitedefaultendpunct}{\mcitedefaultseppunct}\relax
\EndOfBibitem
\bibitem[Raman(1953)]{raman1953molecular}
Raman,~C.~V. The molecular scattering of light. Nobel lecture delivered at
  Stockholm, 11th december 1930. Proc. Indian Acad. Sci. A. 1953; pp
  342--349\relax
\mciteBstWouldAddEndPuncttrue
\mciteSetBstMidEndSepPunct{\mcitedefaultmidpunct}
{\mcitedefaultendpunct}{\mcitedefaultseppunct}\relax
\EndOfBibitem
\bibitem[Ramanathan(1922)]{ramanathan1922molecular}
Ramanathan,~K.~R. \emph{Proc. R. Soc. Lond. A} \textbf{1922}, \emph{102},
  151--161\relax
\mciteBstWouldAddEndPuncttrue
\mciteSetBstMidEndSepPunct{\mcitedefaultmidpunct}
{\mcitedefaultendpunct}{\mcitedefaultseppunct}\relax
\EndOfBibitem
\bibitem[Ramanathan(1923)]{ramanathan1923lviii}
Ramanathan,~K.~R. \emph{Lond.Edinb.Dubl.Phil.Mag.} \textbf{1923}, \emph{46},
  543--553\relax
\mciteBstWouldAddEndPuncttrue
\mciteSetBstMidEndSepPunct{\mcitedefaultmidpunct}
{\mcitedefaultendpunct}{\mcitedefaultseppunct}\relax
\EndOfBibitem
\bibitem[Ramanathan(1925)]{ramanathan1925transparency}
Ramanathan,~K.~R. \emph{Phys. Rev.} \textbf{1925}, \emph{25}, 386\relax
\mciteBstWouldAddEndPuncttrue
\mciteSetBstMidEndSepPunct{\mcitedefaultmidpunct}
{\mcitedefaultendpunct}{\mcitedefaultseppunct}\relax
\EndOfBibitem
\bibitem[Shoulejkin(1923)]{shoulejkin1923color}
Shoulejkin,~W. \emph{Phys. Rev.} \textbf{1923}, \emph{22}, 85\relax
\mciteBstWouldAddEndPuncttrue
\mciteSetBstMidEndSepPunct{\mcitedefaultmidpunct}
{\mcitedefaultendpunct}{\mcitedefaultseppunct}\relax
\EndOfBibitem
\bibitem[Shoulejkin(1924)]{shoulejkin1924color}
Shoulejkin,~W. \emph{Phys. Rev.} \textbf{1924}, \emph{23}, 744\relax
\mciteBstWouldAddEndPuncttrue
\mciteSetBstMidEndSepPunct{\mcitedefaultmidpunct}
{\mcitedefaultendpunct}{\mcitedefaultseppunct}\relax
\EndOfBibitem
\bibitem[Dickey \latin{et~al.}(2011)Dickey, Kattawar, and
  Voss]{dickey2011shedding}
Dickey,~T.~D.; Kattawar,~G.~W.; Voss,~K.~J. \emph{Phys. Today} \textbf{2011},
  \emph{64}, 44--49\relax
\mciteBstWouldAddEndPuncttrue
\mciteSetBstMidEndSepPunct{\mcitedefaultmidpunct}
{\mcitedefaultendpunct}{\mcitedefaultseppunct}\relax
\EndOfBibitem
\bibitem[Jerlov(1976)]{jerlov1976marine}
Jerlov,~N. \emph{Marine Optics}; Elsevier Oceanography Series; Elsevier
  Science, 1976; p 165\relax
\mciteBstWouldAddEndPuncttrue
\mciteSetBstMidEndSepPunct{\mcitedefaultmidpunct}
{\mcitedefaultendpunct}{\mcitedefaultseppunct}\relax
\EndOfBibitem
\bibitem[Shoulejkin(1933)]{shoulejkin1933data}
Shoulejkin,~W. \emph{Geofiz.} \textbf{1933}, \emph{3}, 3--5, cit.
  in~\citenum{jerlov1976marine}, p.~83\relax
\mciteBstWouldAddEndPuncttrue
\mciteSetBstMidEndSepPunct{\mcitedefaultmidpunct}
{\mcitedefaultendpunct}{\mcitedefaultseppunct}\relax
\EndOfBibitem
\bibitem[Wernand(2011)]{phdthesis}
Wernand,~M.~R. Poseidon’s paintbox: historical archives of ocean colour in
  global--change perspective. Ph.D.\ thesis, University of Utrecht, 2011;
  disponibile all'url:
  \url{https://www.researchgate.net/publication/254886155_Poseidon%27s_paintbox_historical_archives_of_ocean_colour_in_global-change_perspective}\relax
\mciteBstWouldAddEndPuncttrue
\mciteSetBstMidEndSepPunct{\mcitedefaultmidpunct}
{\mcitedefaultendpunct}{\mcitedefaultseppunct}\relax
\EndOfBibitem
\bibitem[Wernand and Gieskes(2011)Wernand, and Gieskes]{wernand2011ocean}
Wernand,~M.~R.; Gieskes,~W. W.~C. \emph{Ocean optics from 1600 (Hudson) to 1930
  (Raman), shifting interpretation of natural water colouring}; Union des
  Oc{\'e}anographes de France, Paris, France, 2011\relax
\mciteBstWouldAddEndPuncttrue
\mciteSetBstMidEndSepPunct{\mcitedefaultmidpunct}
{\mcitedefaultendpunct}{\mcitedefaultseppunct}\relax
\EndOfBibitem
\bibitem[Cabannes(1949)]{cabannes1949azul}
Cabannes,~J. \emph{Ciencia e invest.} \textbf{1949}, \emph{5}, 3--17, cit.
  in~\citenum{lenoble1956remarque}\relax
\mciteBstWouldAddEndPuncttrue
\mciteSetBstMidEndSepPunct{\mcitedefaultmidpunct}
{\mcitedefaultendpunct}{\mcitedefaultseppunct}\relax
\EndOfBibitem
\bibitem[Lenoble(1956)]{lenoble1956remarque}
Lenoble,~J. \emph{Comptes Rendus} \textbf{1956}, \emph{242}, 662--664\relax
\mciteBstWouldAddEndPuncttrue
\mciteSetBstMidEndSepPunct{\mcitedefaultmidpunct}
{\mcitedefaultendpunct}{\mcitedefaultseppunct}\relax
\EndOfBibitem
\bibitem[Chandrasekhar(1950)]{chandrasekhar1950radiative}
Chandrasekhar,~S. \emph{Radiative Transfer}; Dover books on physics and
  engineering; Clarendon Press, 1950\relax
\mciteBstWouldAddEndPuncttrue
\mciteSetBstMidEndSepPunct{\mcitedefaultmidpunct}
{\mcitedefaultendpunct}{\mcitedefaultseppunct}\relax
\EndOfBibitem
\bibitem[Le~Grand(1939)]{legrand1939penetr}
Le~Grand,~Y. \emph{Ann. Inst. Oceanogr. (Paris)} \textbf{1939}, \emph{19},
  393--436, cit. in~\citenum{lenoble1956remarque}, p.~662\relax
\mciteBstWouldAddEndPuncttrue
\mciteSetBstMidEndSepPunct{\mcitedefaultmidpunct}
{\mcitedefaultendpunct}{\mcitedefaultseppunct}\relax
\EndOfBibitem
\bibitem[Smith and Baker(1981)Smith, and Baker]{smith1981optical}
Smith,~R.~C.; Baker,~K.~S. \emph{Appl. Opt.} \textbf{1981}, \emph{20},
  177--184\relax
\mciteBstWouldAddEndPuncttrue
\mciteSetBstMidEndSepPunct{\mcitedefaultmidpunct}
{\mcitedefaultendpunct}{\mcitedefaultseppunct}\relax
\EndOfBibitem
\bibitem[Un~approccio di questo tipo è adottato ad es.~in Nielsen and
  Jerlov(1974)Un~approccio di questo tipo è adottato ad es.~in Nielsen, and
  Jerlov]{nielsen1974optical}
Un~approccio di questo tipo è adottato ad es.~in Nielsen,~E.; Jerlov,~N.
  \emph{Optical Aspects of Oceanography: Papers Presented at the Symposium on
  Optical Aspects of Oceanography, Held at the Institute of Physical
  Oceanography in Copenhagen, 19-23 June 1972}; Academic Press, 1974; cit.
  in~\citenum{plass1978color}, p.~1433\relax
\mciteBstWouldAddEndPuncttrue
\mciteSetBstMidEndSepPunct{\mcitedefaultmidpunct}
{\mcitedefaultendpunct}{\mcitedefaultseppunct}\relax
\EndOfBibitem
\bibitem[Gordon and Morel(2012)Gordon, and Morel]{gordon2012remote}
Gordon,~H.; Morel,~A. \emph{Remote Assessment of Ocean Color for Interpretation
  of Satellite Visible Imagery: A Review}; Coastal and Estuarine Studies;
  Springer New York, 2012; Vol.~4\relax
\mciteBstWouldAddEndPuncttrue
\mciteSetBstMidEndSepPunct{\mcitedefaultmidpunct}
{\mcitedefaultendpunct}{\mcitedefaultseppunct}\relax
\EndOfBibitem
\bibitem[Not()]{Note-2}
La letteratura sull'argomento è davvero sterminata. Per un primo riferimento
  si possono esplorare le seguenti pagine di Wikipedia:
  \url{https://en.wikipedia.org/wiki/Electromagnetic_absorption_by_water},
  \url{https://en.wikipedia.org/wiki/Molecular_vibration}.\relax
\mciteBstWouldAddEndPunctfalse
\mciteSetBstMidEndSepPunct{\mcitedefaultmidpunct}
{}{\mcitedefaultseppunct}\relax
\EndOfBibitem
\bibitem[Un'ampia trattazione del soggetto si trova~in Wozniak and
  Dera(2007)Un'ampia trattazione del soggetto si trova~in Wozniak, and
  Dera]{wozniak2007light}
Un'ampia trattazione del soggetto si trova~in Wozniak,~B.; Dera,~J. \emph{Light
  Absorption in Sea Water}; Atmospheric and Oceanographic Sciences Library;
  Springer New York, 2007\relax
\mciteBstWouldAddEndPuncttrue
\mciteSetBstMidEndSepPunct{\mcitedefaultmidpunct}
{\mcitedefaultendpunct}{\mcitedefaultseppunct}\relax
\EndOfBibitem
\bibitem[per una rassegna degli sviluppi teorici e sperimentali per lo spettro
  di assorbimento dell'acqua nella fase di vapore~v. Bernath(2002)]{B200372D}
per una rassegna degli sviluppi teorici e sperimentali per lo spettro di
  assorbimento dell'acqua nella fase di vapore~v. Bernath,~P.~F. \emph{Phys.
  Chem. Chem. Phys.} \textbf{2002}, \emph{4}, 1501--1509\relax
\mciteBstWouldAddEndPuncttrue
\mciteSetBstMidEndSepPunct{\mcitedefaultmidpunct}
{\mcitedefaultendpunct}{\mcitedefaultseppunct}\relax
\EndOfBibitem
\bibitem[sull'origine vibrazionale del colore dell'acqua v.~anche Braun and
  Smirnov(1993)sull'origine vibrazionale del colore dell'acqua v.~anche Braun,
  and Smirnov]{braun1993water}
sull'origine vibrazionale del colore dell'acqua v.~anche Braun,~C.~L.;
  Smirnov,~S.~N. \emph{J. Chem. Educ.} \textbf{1993}, \emph{70}, 612--612\relax
\mciteBstWouldAddEndPuncttrue
\mciteSetBstMidEndSepPunct{\mcitedefaultmidpunct}
{\mcitedefaultendpunct}{\mcitedefaultseppunct}\relax
\EndOfBibitem
\bibitem[Per una trattazione introduttiva della composizione chimica del
  mare~v. Pilson(2013)]{pilson2013introduction}
Per una trattazione introduttiva della composizione chimica del mare~v.
  Pilson,~M. \emph{An Introduction to the Chemistry of the Sea}; Cambridge
  University Press, 2013\relax
\mciteBstWouldAddEndPuncttrue
\mciteSetBstMidEndSepPunct{\mcitedefaultmidpunct}
{\mcitedefaultendpunct}{\mcitedefaultseppunct}\relax
\EndOfBibitem
\bibitem[Young(1981)]{young1981rayleigh}
Young,~A.~T. \emph{Appl. Opt.} \textbf{1981}, \emph{20}, 533--535\relax
\mciteBstWouldAddEndPuncttrue
\mciteSetBstMidEndSepPunct{\mcitedefaultmidpunct}
{\mcitedefaultendpunct}{\mcitedefaultseppunct}\relax
\EndOfBibitem
\bibitem[Young(1982)]{young1982rayleigh}
Young,~A.~T. \emph{Phys. Today} \textbf{1982}, \emph{35}, 42--48\relax
\mciteBstWouldAddEndPuncttrue
\mciteSetBstMidEndSepPunct{\mcitedefaultmidpunct}
{\mcitedefaultendpunct}{\mcitedefaultseppunct}\relax
\EndOfBibitem
\bibitem[Mandel'shtam(1926)]{mandelstam1926light}
Mandel'shtam,~L.~I. \emph{Zh. Russ. Fiz-Khim. Ova} \textbf{1926}, \emph{58},
  381\relax
\mciteBstWouldAddEndPuncttrue
\mciteSetBstMidEndSepPunct{\mcitedefaultmidpunct}
{\mcitedefaultendpunct}{\mcitedefaultseppunct}\relax
\EndOfBibitem
\bibitem[Brillouin(1922)]{brillouin1922diffusion}
Brillouin,~L. Diffusion de la lumi{\`e}re et des rayons X par un corps
  transparent homog{\`e}ne-Influence de l'agitation thermique. Ann. de Phys.
  1922; pp 88--122\relax
\mciteBstWouldAddEndPuncttrue
\mciteSetBstMidEndSepPunct{\mcitedefaultmidpunct}
{\mcitedefaultendpunct}{\mcitedefaultseppunct}\relax
\EndOfBibitem
\bibitem[Fabelinskii(2012)]{fabelinskii2012molecular}
Fabelinskii,~I.~L. \emph{Molecular Scattering of Light}; Springer US,
  2012\relax
\mciteBstWouldAddEndPuncttrue
\mciteSetBstMidEndSepPunct{\mcitedefaultmidpunct}
{\mcitedefaultendpunct}{\mcitedefaultseppunct}\relax
\EndOfBibitem
\bibitem[Gross(1930)]{gross1930change}
Gross,~E. \emph{Nature} \textbf{1930}, \emph{126}, 201\relax
\mciteBstWouldAddEndPuncttrue
\mciteSetBstMidEndSepPunct{\mcitedefaultmidpunct}
{\mcitedefaultendpunct}{\mcitedefaultseppunct}\relax
\EndOfBibitem
\bibitem[Landau and Placzek(1934)Landau, and Placzek]{landau1934structure}
Landau,~L.; Placzek,~G. \emph{Phys. Z. Sowiet. Un.} \textbf{1934}, \emph{5},
  172\relax
\mciteBstWouldAddEndPuncttrue
\mciteSetBstMidEndSepPunct{\mcitedefaultmidpunct}
{\mcitedefaultendpunct}{\mcitedefaultseppunct}\relax
\EndOfBibitem
\bibitem[Chen \latin{et~al.}(2018)Chen, Zhang, and Guan]{CHEN201843}
Chen,~W.; Zhang,~T.; Guan,~L. In \emph{Comprehensive Remote Sensing};
  Liang,~S., Ed.; Elsevier: Oxford, 2018; pp 43 -- 78\relax
\mciteBstWouldAddEndPuncttrue
\mciteSetBstMidEndSepPunct{\mcitedefaultmidpunct}
{\mcitedefaultendpunct}{\mcitedefaultseppunct}\relax
\EndOfBibitem
\bibitem[Adam(2011)]{adam2011mathematical}
Adam,~J.~A. \emph{A Mathematical Nature Walk}; Princeton University Press,
  2011; pp 77--79\relax
\mciteBstWouldAddEndPuncttrue
\mciteSetBstMidEndSepPunct{\mcitedefaultmidpunct}
{\mcitedefaultendpunct}{\mcitedefaultseppunct}\relax
\EndOfBibitem
\bibitem[Freiherr~v.u.z. Aufsess(2013)]{von2013physikalischen}
Freiherr~v.u.z. Aufsess,~O. \emph{Die Physikalischen Eigenschaften der Seen};
  Die Wissenschaft; Vieweg+Teubner Verlag, 2013\relax
\mciteBstWouldAddEndPuncttrue
\mciteSetBstMidEndSepPunct{\mcitedefaultmidpunct}
{\mcitedefaultendpunct}{\mcitedefaultseppunct}\relax
\EndOfBibitem
\bibitem[Freiherr~v.u.z. Aufsess(1904)]{frhr1904farbe}
Freiherr~v.u.z. Aufsess,~O. \emph{Ann. Phys. (Berlin)} \textbf{1904},
  \emph{318}, 678--711\relax
\mciteBstWouldAddEndPuncttrue
\mciteSetBstMidEndSepPunct{\mcitedefaultmidpunct}
{\mcitedefaultendpunct}{\mcitedefaultseppunct}\relax
\EndOfBibitem
\bibitem[Mees(1909)]{mees1909atlas}
Mees,~C. E.~K. \emph{Atlas of absorption spectra}; Longmans, Green, 1909\relax
\mciteBstWouldAddEndPuncttrue
\mciteSetBstMidEndSepPunct{\mcitedefaultmidpunct}
{\mcitedefaultendpunct}{\mcitedefaultseppunct}\relax
\EndOfBibitem
\bibitem[Not()]{Note-3}
Approfondimenti sulla questione si possono trovare al link:
  \url{http://www1.lsbu.ac.uk/water/water_sitemap.html}; un'ottima risorsa è
  anche l'\texttt{Optical Absorption of Water Compendium} al link:
  \url{https://omlc.org/spectra/water/abs/index.html}\relax
\mciteBstWouldAddEndPuncttrue
\mciteSetBstMidEndSepPunct{\mcitedefaultmidpunct}
{\mcitedefaultendpunct}{\mcitedefaultseppunct}\relax
\EndOfBibitem
\end{mcitethebibliography}

\end{document}